\documentclass{article}

\usepackage{xcolor}
\definecolor{tabblue}{rgb}{0.133, 0.475, 0.706}
\definecolor{taborange}{rgb}{1.000, 0.490, 0.047}
\definecolor{tabred}{rgb}{0.839, 0.168, 0.168}
\definecolor{tabgreen}{rgb}{0.169, 0.620, 0.169}

\usepackage[section]{placeins}
\usepackage{float}
\usepackage[title]{appendix}
\usepackage{amsmath} 
\usepackage{amssymb}
\usepackage{textcomp}
\usepackage{bm}
\usepackage{graphicx}
\usepackage{graphics}
\usepackage{subcaption}
\usepackage[symbol]{footmisc}

\usepackage{PRIMEarxiv}

\usepackage[utf8]{inputenc} 
\usepackage[T1]{fontenc}    
\usepackage{hyperref}       
\usepackage{url}            
\usepackage{booktabs}       
\usepackage{amsfonts}       
\usepackage{nicefrac}       
\usepackage{microtype}      
\usepackage{lipsum}
\usepackage{fancyhdr}       
\usepackage{graphicx}       
\graphicspath{{media/}}     

\title{A Bayesian model calibration framework for stochastic compartmental models with both time-varying and time-invariant parameters
}

\author{
  Brandon Robinson\\
  Department of Civil and Environmental Engineering\\
  Carleton University \\
  Ottawa, ON, Canada\\
   \And
  Philippe Bisaillon \\
  Department of Civil and Environmental Engineering\\
  Carleton University \\
  Ottawa, ON, Canada \\
   \And
  Jodi D. Edwards\\
  School of Epidemiology and Public Health \\ 
  University of Ottawa \\
 and University of Ottawa Heart Institute\\
  Ottawa, ON, Canada\\
   \And
  Tetyana Kendzerska\\
 Department of Medicine, Faculty of Medicine, Division of Respirology \\ University of Ottawa \\  
 and The Ottawa Hospital Research Institute\\
  Carleton University \\
  Ottawa, ON, Canada\\
   \And
  Mohammad Khalil
\thanks{\textit{Sandia National Laboratories is a multimission laboratory managed and operated by National Technology and Engineering Solutions of Sandia, LLC., a wholly owned subsidiary of Honeywell International, Inc., for the U.S. Department of Energy’s National Nuclear Security Administration under contract DE-NA-0003525.}} \\ 
  Quantitative Modeling \& Analysis Department\\
  Sandia National Laboratories \\
  Livermore, CA, United States\\
   \And
  Dominique Poirel \\
  Department of Mechanical and Aerospace Engineering\\
  Royal Military College of Canada \\
  Kingston, ON, Canada \\
   \And
  Abhijit Sarkar \\
  Department of Civil and Environmental Engineering\\
  Carleton University \\
  Ottawa, ON, Canada\\
}

\begin{document}
\maketitle
\begin{abstract}
We consider state and parameter estimation for compartmental models having both time-varying and time-invariant parameters. Though the described Bayesian computational framework is general, we look at a specific application to the susceptible-infectious-removed (SIR) model which describes a basic mechanism for the spread of infectious diseases through a system of coupled nonlinear differential equations. The SIR model consists of three states, namely, the three compartments, and two parameters which control the coupling among the states. The deterministic SIR model with time-invariant parameters has shown to be overly simplistic for modelling the complex long-term dynamics of diseases transmission. Recognizing that certain model parameters will naturally vary in time due to seasonal trends, non-pharmaceutical interventions, and other random effects, the estimation procedure must systematically permit these time-varying effects to be captured, without unduly introducing artificial dynamics into the system. To this end, we leverage the robustness of the Markov Chain Monte Carlo (MCMC) algorithm for the estimation of time-invariant parameters alongside nonlinear filters for the joint estimation of the system state and time-varying parameters. We demonstrate performance of the framework by first considering a series of examples using synthetic data, followed by an exposition on public health data collected in the province of Ontario.
\end{abstract}

\section{Introduction}
This paper presents an algorithm for the combined state and parameter estimation of stochastic compartmental models having both time-varying and time-invariant parameters. The methodology described herein was originally motivated by system identification in nonlinear dynamical systems with a specific interest in mechanical systems \cite{khalil2013probabilistic,khalil2015estimation,bisaillon2015bayesian}.  Bisaillon et al. \cite{bisaillon2022robust} describe a fully Bayesian model selection framework which leverages the use of nonlinear filters and Markov Chain Monte Carlo (MCMC) methods for the combined state and parameter estimation of dynamical systems with both time-varying and time-invariant parameters. Even though the methodology is inspired by mechanical systems, it operates on general nonlinear dynamical systems, which naturally lends itself to infectious disease modelling using compartmental models. It is particularly relevant due to the fact that many system parameters in a compartmental model may change over the course of a given epidemic or pandemic, whereas others will tend to remain constant (time-invariant). The complimentary use of nonlinear filters and MCMC provides a robust computational framework which systematically alleviates many limitations of traditional methods for time-varying parameter estimation using just nonlinear filters. 

There are many examples of the Kalman filter and other nonlinear filters being used for tracking case counts
\cite{zeng2020dynamics,arroyo2021tracking,singh2021kalman,sun2023analysis,evensen2021international}. When accounting for model discrepancy using stochastic compartmental models, Mamis \& Farazmand \cite{mamis2023stochastic} suggested that the use of correlated noise should be favoured over white noise. Notably, the mechanism by which we estimate the time-varying system parameters are mathematically analogous to a stochastic compartmental model with a coloured noise process (as explained later). Traditional methods for combined state and parameter estimation using nonlinear filters operate by augmenting the state vector to include the system parameters. By appending the time-varying parameters to the state vector, these parameters can be jointly estimated alongside the system states \cite{shumway2000time,evensen2009data}. In order to ensure filter convergence, the parameters are perturbed by an artificial random noise, ensuring that some uncertainty is maintained in the parameter estimates as time progresses. While this  artificial noise resolves the issue of filter divergence, it introduces three new points of concern. First, by treating the system parameters as states in the augmented system, the degree of nonlinearity in the state space model is increased, which in practice may limit the modeller's ability to use certain filters \cite{khalil2015estimation}. Second, by permitting both time-varying and time-invariant parameters to vary in time for the purpose of estimation, artificial (or non-physical) dynamics are introduced into the system. Third, by perturbing the parameters by an artificial random noise process, the covariance of this system noise contains additional parameters that must be tuned or estimated in a robust fashion.

The Bayesian computational framework described here alleviates each of these concerns through the combination of sampling methods and nonlinear filters. Relying on MCMC for the time-invariant parameter estimation means that the state vector need only be augmented to include the original system states and time-varying parameters. By only appending the subset of parameters that are known to (or assumed to) vary in time to the state vector, the degree of nonlinearity in the augmented state space model will typically be reduced vis-a-vis the case where all parameters are modelled in this fashion. Furthermore, the parameters that are known to be time-invariant behave as such within the state estimation procedure. Hence, the strength of the artificial dynamics introduced by the estimation procedure is reduced. Finally, the question of estimating the process noise covariance matrix is critically addressed by this setup. By treating the parameters of the covariance matrix as time-invariant parameters to be jointly estimated alongside the set of time-invariant system parameters by MCMC, we provide a robust, fully Bayesian approach to estimate the artificial noise strength as dictated by the system dynamics and data. The caveat of this approach is the increased computational overhead associated with MCMC. The time-scale of infectious disease spread and the frequency of data collection is on the order of days compared to mechanical systems which may be on the order of seconds. Hence, for infectious disease modelling, where uncertainties in the states and parameters are abundant, the increased time required for computation is justified.

In addition to addressing some shortcomings of existing methods for combined state and parameter estimation using filtering methods, this framework critically allows us to directly address some major criticisms of the use of SIR models for long-term predictions of infectious disease spread. Some researchers have discussed the inadequacy of the SIR model to capture the complexity of a global phenomenon like COVID-19 \cite{moein2021inefficiency,ellison2020implications}, whereas others have discussed the pitfalls of more complex models regarding the identifiability of parameters and the observability of additional compartments \cite{massonis2021structural,gallo2022lack}. These studies essentially point out that the complexity of the model and availability of data should be considered hand-in-hand to select optimal models capable of meaningful predictions. For aggregated population-level modelling, the inclusion of a time-varying infection rate parameter, estimated within a robust fully Bayesian computational framework addresses these sources of concern concurrently. Note that the definition of a meaningful prediction itself is predicated on the desired spatial and temporal scales, sources of uncertainty and other modelling assumptions \cite{holmdahl2020wrong}. Where the simplicity of a deterministic SIR model with constant parameters is mathematically incapable of generating multiple waves of infection, an SIR model augmented by a time-varying infection rate permits such dynamics to be captured. Furthermore, in addition to the time-varying nature of the disease transmission, other sources of model discrepancy are also captured through this parameter, which is akin to a model error term \cite{evensen2009data}. Another prominent issue impacting our understanding of the course of the pandemic and our ability to forecast the future trajectory of case counts is the unknown portion of cases that are undetected despite widespread testing efforts \cite{mukhopadhyay2020estimation,shaman2021estimation,rippinger2021evaluation}. Such predictions are critical to informing policy decisions regarding interventions and healthcare resource allocations based on predicted case counts and hospitalizations \cite{zhang2022learning,squire2021modeling,kendzerska2022trends}. To address these concerns in modelling efforts, some researchers have attempted to systematically address the under reporting of cases by introducing explicit compartments that differentiate between detected and undetected cases \cite{gaeta2020simple,liu2022modeling,olumoyin2021data}. In our case, we have chosen to parameterize the measurement operator, introducing a multiplication factor to be estimated which quantifies the proportion of active cases are actually reported. While this multiplication factor is treated as a time-invariant parameter in the current investigation, it could be treated as a time-varying parameter within the framework to capture potential fluctuations in testing rates.

Section \ref{sec:problemstatement} outlines the mathematical methods employed in this study. The SIR model and its discrete state space form are described in Section \ref{sec:sir}. The Bayesian computational framework and its implementation are described in Section \ref{sec:framework}. The results are presented in Section \ref{sec:results}. Section \ref{sec:synthetic} considers model calibration using a series of synthetic datasets that mimic observed trends in public health data collected and reported at the population level during the COVID-19 pandemic. Using known time-invariant parameters and defining three unique time-varying infection rate parameters representing i) random variations in the infection rate, ii) gradual seasonal variations in the infection rate, modelled by a sinusoidal function, and iii) a sudden and significant change in the infection rate to simulate a lockdown, modelled by a sigmoid function. These cases permits the assessment of the algorithmic performance vis-a-vis its ability to recover the ground truth parameter values from which the data were generated. We then apply the same SIR model and inference setup to publicly reported active case data \cite{OntarioI} from the province of Ontario as a case study in Section \ref{sec:ontario}.

\section{Material and methods}\label{sec:problemstatement}
This section contains a description introducing the compartmental models used in this study, and the available public health data that will be essential to the estimation of time-varying and time-invariant parameters. Subsequently, the Bayesian algorithm for parameter estimation is described for a general state-space model.

\subsection{SIR Model}\label{sec:sir}
\begin{figure}[!htb]
\centering
\includegraphics[width=\textwidth]{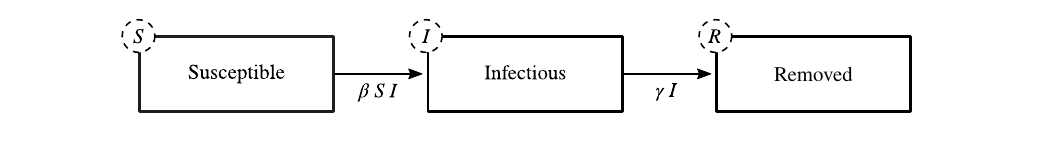}
\caption{Three-compartment SIR model representing Eqs. (\ref{eq:3compartment1} - \ref{eq:3compartment3}).}
\label{fig:compartments}
\end{figure}

The SIR model consists of three compartments, namely, susceptible ($S$), infectious ($I$), and removed ($R$), also commonly referred to as recovered or resolved. The three compartments classify the population based on their current disease state. The relationship between the three compartments, and dynamics of how the population interacts with the disease in question can be represented by the flowchart in Figure \ref{fig:compartments}. Mathematically, the dynamics are described by the following system of first-order coupled nonlinear ordinary differential equations (ODEs)

\begin{align*}
\frac{dS}{dt} &= -\beta S(t) I(t) \stepcounter{equation}\tag{{\theequation}a} \label{eq:3compartment1} ,\\
\frac{dI}{dt} &= \beta S(t) I(t) - \gamma I(t) \tag{{\theequation}b}\label{eq:3compartment2}, \\
\frac{dR}{dt} &=  \gamma I(t) \tag{{\theequation}c}\label{eq:3compartment3}.
\end{align*} 
where the compartments $S$, $I$, and $R$ are model states, and $\beta$ and $\gamma$ are the system parameters. This system of coupled ODEs can be rewritten as a discrete state space model, which can be written as the following system of algebraic equations when using the forward Euler method,


\begin{subequations}
\begin{align}
S_{k+1} &= S_{k} + \Delta t \left( -\beta S_{k}I_{k} \right), 
\\
I_{k+1} &= I_{k} + \Delta t \left( \beta S_{k}I_{k} - \gamma I_{k} \right), 
\\
R_{k+1} &= R_{k} + \Delta t \left( \gamma I_{k} \right),
\end{align}
\end{subequations}
the system states at discrete time instance $t_k$ along the computational grid are $S_k$, $I_k$, and $R_k$, $\Delta t$ is the timestep.

The susceptible compartment consists of all individuals that are vulnerable to becoming infected with the disease. The infectious compartment accounts for all the active infections. An individual will move from the susceptible compartment to the infectious compartment if and only if they engage in a disease transmitting interaction with an individual who is infectious. This is represented by the quadratic nonlinear term $\beta S I$, where the parameter $\beta$ reflects the average number of daily interactions an individual is expected to have, and the probability that any one of those interactions between a susceptible and an infectious individual will result in the disease being transmitted. The parameter $\beta$, may vary in time as a result of the implementation of non-pharmaceutical interventions, such as lockdown measures which reduces the number of daily interactions, as well as a result of various public health measures such as mask mandates or vaccination, which reduce the probability that a given interaction will result in the transmission of the disease. The removed compartment in this model is a terminal compartment, as we do not consider reinfection. Furthermore, we account for both recovered and dead individuals within the removed compartment. These two aspects can are generalized in other basic compartmental such as the SIRS model 
(e.g., \cite{hu2020heterogeneity,hooten2010assessing}) which allows for reinfection after a period of temporary immunity, and the SIRD model (e.g., \cite{calafiore2020time,fernandez2022estimating}) that has two separate terminal compartments differentiating recovered and deceased individuals. In any event, the removal of individuals from the infectious compartment is controlled by the term $\gamma I$, wherein the parameter $\gamma$ represents a rate equal to the inverse of the average duration of infection. This parameter, $\gamma$, will tend to be constant in time. However, for certain diseases, it may still vary due to new strains or variants of a virus emerging, or as treatments become available. For the sufficiently short time-horizon considered in this study, the parameter $\gamma$ can reasonably be taken to be time-invariant.

The SIR model may become stochastic by introducing of a model error term. The model error is implemented systematically to ensure conservation \cite{tornatore2005stability,zhang2013stochastic,ji2014threshold}. This model error permits unmodelled disease dynamics to be accounted for by the stochastic volatility of the parameters. Alternatively, we would like to model the parameter $\beta$ as a time-varying quantity allowing the volatility in the population-wide disease dynamics to be reflected by recursively estimating it. To permit the estimation of the time-varying parameter $\beta$, we append it to the state vector, augmenting the dimension of the state space and introducing an additional coupled differential equation to the system in Eq. (\ref{eq:3compartment1} - \ref{eq:3compartment3}),

\begin{equation}\label{eq:randombeta}
\frac{d \beta}{dt} = q_\xi \xi(t)
\end{equation}
where this rate of change of the time-varying parameter $\beta$, is described by a white noise Gaussian process, thereby resulting.in the following discretized system of equations,
 
\begin{subequations}\label{eq:sirb}
\begin{align}
S_{k+1} &= S_{k} + \Delta t \left( -\beta_{k} S_{k}I_{k} \right), 
\\
I_{k+1} &= I_{k} + \Delta t \left( \beta_{k} S_{k}I_{k} - \gamma I_{k} \right), 
\\
R_{k+1} &= R_{k} + \Delta t \left( \gamma I_{k} \right),
\\
\beta_{k+1} &= \beta_{k} + q_\xi \sqrt{\Delta t}  \varepsilon_{k}.
\end{align}
\end{subequations}
where $\varepsilon_k$ is a zero mean Gaussian random variable with unit variance. The parameter $q_\xi$ controls the strength of the artificial noise and is an additional static parameter to be estimated. In the form of a general nonlinear state space model, we can summarize Eqs. (\ref{eq:sirb}) as

\begin{equation}
\mathbf{x}_{k+1} = \mathbf{g}_k(\mathbf{x}_k, \bm{\Phi}_s,\mathbf{q}_k).\label{eq:model_op}
\end{equation}
where $\mathbf{g}_k$ is the nonlinear model operator, $\mathbf{x}_k = \{x_{1,k},x_{2,k},x_{3,k},x_{4,k}\}^T = \{S_k,I_k,R_k,\beta_k\}^T$ is the augmented state vector, $\bm{\Phi}_s$ is the set of time-invariant parameters, and $\mathbf{q}_k = \{0,0,0,q_\xi\}^T$ is the stochastic forcing.

%


All of the parameters are to be estimated using aggregated public health data, reporting daily public testing results for COVID-19. The Bayesian framework used herein provides a robust approach for blending measurements with model predictions. The public health data is characterized by the following general measurement equation ~\cite{evensen2009data,jazwinski2007stochastic}

\begin{equation}
 \mathbf{d}_{j} = \mathbf{h}_{j} (\mathbf{x}_{d(j)},\bm{\Phi}_s, \mathbf{\epsilon}_{j}) \label{eq:meas}
\end{equation}
where $\mathbf{d}_{j}$ is the noisy observation of the state vector at time step $t_{d(j)}$ and $\mathbf{\epsilon}_j$ is the measurement noise modelled as independent zero-mean Gaussian random vector with a known covariance matrix. For the current investigation, the measurement operator defines a multiplicative Gaussian noise, with a parameter $\rho$, which characterizes the observed data as a partial account of the true number of infected individuals. The noise in the data accounts for false positives and negatives, while the parameter $\rho$ accounts for the fact that the number of recorded positive tests do not reflect a complete observation of the entire population at all time instance $t_{d(j)}$.

\begin{figure}[!htb]
\centering
\includegraphics[width=0.8\linewidth,keepaspectratio]{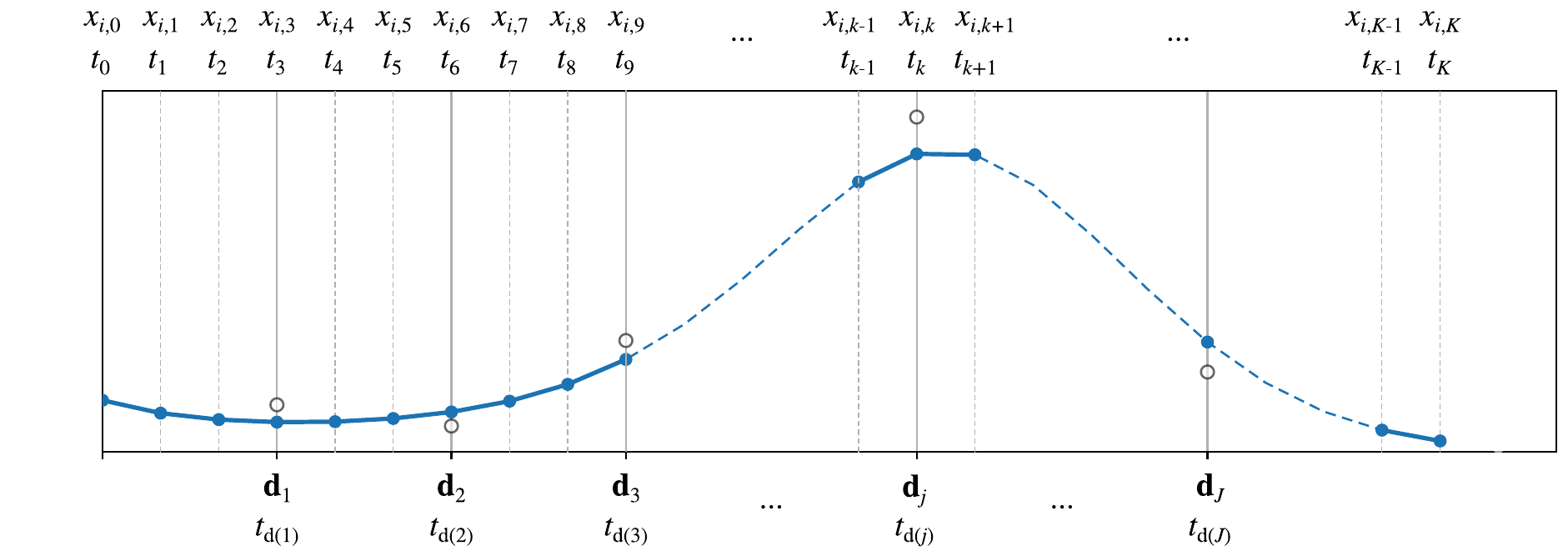}
\caption{The computational grid in Eq.~(\ref{eq:model_op}) is indexed by $k$, depicted by filled circles ($\textcolor{tabblue}{\bullet}$). The data in Eq.~(\ref{eq:meas}) are indexed by $j$, depicted by empty circles ($\circ$). }
\label{fig:indices}
\end{figure}

The simplicity of the SIR model permits the convenient computation of the effective reproduction number. The effective reproduction number is a critical metric which may be interpreted as the average number of new infections that will result from each active infection. This is conveniently computed as a function of the two system parameters \cite{brauer2008mathematical},

\begin{equation}
R_e(t) = \frac{S(t)\beta(t)}{\gamma}. \label{eq:Re}
\end{equation} 
where $S(t) \beta (t)$ indicates the daily number of disease transmitting interactions a given infectious individual will have with susceptible individuals, and $1/ \gamma$ is the average number of days that an individual will be infectious. This metric is a generalized form of the basic reproduction number, where a value of $R_e(t) \geq 1$ indicates an ongoing outbreak \cite{brauer2008mathematical}.

\subsection{Framework for estimating time-varying and time-invariant parameters}\label{sec:framework}

If we consider an augmented state $\mathbf{x}_k = \{\mathbf{u}_k, \bm{\Phi}_{t(k)}\}$, the joint posterior of the system state, time-invariant and time-varying parameters of model $\mathcal{M}$, conditioned on data $\mathbf{D} = \{\mathbf{d}_0, \hdots,\mathbf{d}_J\}$ can be written as \cite{bisaillon2022robust},

\begin{equation}
\text{p}(\mathbf{u}_1,...,\mathbf{u}_k,\bm{\Phi}_{t(0)},\hdots,\bm{\Phi}_{t(k)}, \bm{\Phi}_s | \mathbf{D}, \mathcal{M})  = \text{p}(\mathbf{x}_1,...,\mathbf{x}_k, \bm{\Phi}_s | \mathbf{D}, \mathcal{M}).
\end{equation}

This joint posterior pdf can then be decomposed by conditioning the augmented state on the static parameters as
\begin{equation}
\text{p}(\mathbf{x}_1,...,\mathbf{x}_k, \bm{\Phi}_s | \mathbf{D}, \mathcal{M}) = \text{p}(\mathbf{x}_1,...,\mathbf{x}_k |\bm{\Phi}_s,  \mathbf{D}, \mathcal{M}) \text{p}(\bm{\Phi}_s | \mathbf{D},\mathcal{M}).
\end{equation}
where $\text{p}(\mathbf{x}_1,...,\mathbf{x}_k |\bm{\Phi}_s,  \mathbf{D}, \mathcal{M})$ is the conditional posterior pdf of the augmented state and $\text{p}(\bm{\Phi}_s | \mathbf{D},\mathcal{M})$ is the posterior pdf of the time-invariant parameters. The computation of the former involves the use of state estimation, while the latter is obtained using MCMC. For a given joint sample of the static parameters, the state estimation procedure using nonlinear filters permits the forecasting of uncertainty in the augmented state in time between measurements and the updating of the state estimates where the computational grid $k$ and measurement $d(j)$ coincide. The state estimation procedure is briefly explained in Appendix \ref{sec:appendixA} and \ref{sec:appendixB}, with details specific to the current implementation using the Extended Kalman Filter.

The time-invariant parameter posterior pdfs are obtained using MCMC. In the current implementation, we use the Transitional MCMC (TMCMC) algorithm \cite{ching2007transitional} to generate samples from the unnormalized time-invariant parameter posterior,

\begin{equation}
\text{p}(\bm{\Phi}_s | \mathbf{D}, \mathcal{M}) = \frac{\text{p}(\mathbf{D} | \bm{\Phi}_s, \mathcal{M}) \text{p}( \bm{\Phi}_s | \mathcal{M})}{\text{p}(\mathbf{D} | \mathcal{M})} \propto \text{p}(\mathbf{D} | \bm{\Phi}_s, \mathcal{M}) \text{p}( \bm{\Phi}_s | \mathcal{M}) ,\label{eq:bayesmodel}
\end{equation}
where $ \text{p}(\bm{\Phi}_s|\mathcal{M})$ is the parameter prior distribution, and $\text{p}(\mathbf{D} |\bm{\Phi}_s,\mathcal{M})$ is the likelihood function. For the purposes of the current investigations, the model evidence $\text{p}(\mathbf{D} |\mathcal{M})$ is simply a normalization factor. Priors may be assigned to the time-invariant parameters in accordance with some known information from clinical studies, or may simply be used to enforce the parameters in the SIR model to be non-negative. For independent measurement noise, the likelihood function in Eq. (\ref{eq:bayesmodel}) is evaluated as the product of the likelihood computed at each data point,

\begin{equation}
\text{p}(\mathbf{D} | \bm{\Phi}_s, \mathcal{M}) = \prod_{j=1}^J \text{p}(\mathbf{d}_j | \bm{\Phi}_s, \mathcal{M}), \label{eq:bayesmodel}
\end{equation}
with

\begin{align}
\text{p}(\mathbf{d}_j | \bm{\Phi}_s, \mathcal{M}) = \int_{-\infty}^{\ \infty} \text{p}( \mathbf{x}_{d(j)} | \mathbf{x}_{d(j)-1}, \bm{\Phi}_s) \text{p}(\mathbf{d}_j | \mathbf{x}_{d(j)} , \bm{\Phi}_s) \text{d} \mathbf{x}_{d(j)}.\label{eq:lik}
\end{align}
which necessitates the use of state estimation \cite{khalil2015estimation,bisaillon2015bayesian}. The state estimation can be performed using Bayesian recursive filters, and expressions for the likelihood function in Eq.~(\ref{eq:lik}) are available in \cite{bisaillon2015bayesian} for various choices of filters. 


\section{Results and Discussion}\label{sec:results}
To demonstrate the performance of this Bayesian computational framework, we consider examples using both synthetic data in Section \ref{sec:synthetic} followed by public health data in Section \ref{sec:ontario}.

For all cases, the measurement operator will be derived from daily testing data, which are partial/incomplete measurements of the infectious compartment, $I$, at time index $k = d(j)$ (i.e. $i_k$ or $x_{2,k}$), corrupted by multiplicative noise. This ensures that the noise in the measurements is higher when case counts are higher. Proportioning the noise strength to the number of reported cases is a more intuitive model compared to additive noise, for instance. Here the noisy and incomplete measurements of the infectious compartment are given by

\begin{equation}
\mathbf{d}_j = \rho i_{d(j)} (1 + \epsilon_j), \qquad \epsilon_j \sim \mathcal{N}(0,{q_\epsilon}^2), \qquad j = 1, \hdots, J
\end{equation}

Using synthetic data representing a sparse, noisy, and partial observation of the state allows us to demonstrate the algorithm's ability to concurrently estimate the model parameters and model states (observed and unobserved). The synthetic data is generated using known parameter values for the time-invariant parameters, and using known functions of time for time-varying parameters. Hence, we are able to test the ability of the algorithm to recover the underlying true parameters and states. We generate three sets of synthetic data, which result from i) a randomly varying infection rate in Section \ref{synthetic_random} modelled by a realization of a Wiener process (the integral of a white noise Gaussian process) 
\cite{kloeden1992stochastic}, ii) a seasonally varying infection rate in Section \ref{synthetic_seasonal} modelled by a sinusoidal function with a period of one year, and iii) a simulated lockdown in Section \ref{synthetic_lockdown} modelled by the sum of two sigmoid functions representing the effect of the sudden enforcement of a lockdown followed by the gradual relaxation of restrictions as observed through monitoring of human mobility patterns during the COVID-19 pandemic \cite{linka2020outbreak,grantz2020use,long2022associations}. Note that in the case where the infection rate parameter happens to be time-invariant, the generality of the framework permits the accurate estimation of this parameter through the state estimation procedure, though these results are omitted for brevity.

The public health data consists of COVID-19 testing data recorded over a period of one year in the province of Ontario \cite{OntarioI}. This real data will inherently embody a combination of the three aforementioned effects that were tested using the synthetic data. It is crucial that the joint estimation of the state, time-varying parameter, and time-invariant parameters be successful when applied to real data, with the caveat that the results cannot be validated as easily as with synthetic data, as the true values of the states and parameters are ultimately unknown.

For the estimation of the time-invariant parameters, we assign parameter prior distributions within the Bayesian framework as listed in Table \ref{table:priors}. Each of the parameters have a direct physical interpretation, and this prior pdf allows us to impose restrictions on the parameter values. We do not favour any specific value for the ratio of observed infections to active infections, $\rho$ and instead use a uniform distribution to simply enforce a lower bound of a 0\% detection rate ($\rho =0$), and an upper bound of 100\% ($\rho =1$) with all values within that range being equally likely a priori. The artificial noise strength $q_\xi$ that perturbs the time-varying infection rate parameter $\beta$, and the estimate of the initial mean of $\beta$ must both be non-negative. Hence, we enforce a lower bound of zero through a uniform distribution for these parameters. The upper bound of the uniform distribution is set to 1 for both parameters as it provides broad support for the parameters given their respective scales. Finally, the initial mean of the infectious compartment is assigned a uniform distribution enforcing that the proportion of individuals initially in the $I$ compartment must be no fewer than 0\%, and no more than 100\%. The average rate of recovery parameter $\gamma$ is taken as a known value of 1/14, and the multiplicative noise strength $q_\epsilon$ is taken as 5\%. The noise strength is estimated based on the deviation of the reported daily cases from the weekly averages in public health data from Ontario \cite{OntarioI}. Considering a trailing weekly average, the noise strength was as high as 9\%, whereas considering a midpoint estimate results in a noise strength closer to 2\%, hence we have selected a value intermediate to these two values. Table \ref{table:priors} also indicates the rate of recovery parameter and the measurement noise strength, which are taken as known and are not estimated here. This choice was based on the reporting of active cases in the real data, which are predicated on an assumed infectious period of 14 days. An active case that does not result in active hospitalization or death is considered to be resolved 14 days after the reported date of symptom onset date. This assumption is maintained in the synthetic noise case for consistency. 

\begin{table}[!h]
\centering
\begin{tabular}{lll}
\hline
Parameter & Symbol  & Prior \\
\hline		
Ratio of observed infections to active infections & $\rho$&  $\mathcal{U}(0,1)$	 \\	
Perturbation strength in parameter $\beta$  & $q_\xi$  & $\mathcal{U}(0,1)$  \\
Initial mean estimate of parameter $\beta$ & $\mathbb{E}[\beta_0]$ &  $\mathcal{U}(0,1)$  \\
Initial mean estimate of state $I$ & $\mathbb{E}[I_0]$ & $\mathcal{U}(0,1)$  \\
\hline							
Rate of recovery & $\gamma$& 1/14	 \\	
Measurement noise strength  & $q_\epsilon$  & 0.05\\
\hline \hline		
\end{tabular}
\caption{Parameter prior pdfs of uncertain parameters.}
\label{table:priors}
\end{table}

\subsection{Synthetic data}\label{sec:synthetic}
\begin{figure}[H]
\centering
\begin{subfigure}{0.33\linewidth}
\includegraphics[width=\linewidth,keepaspectratio]{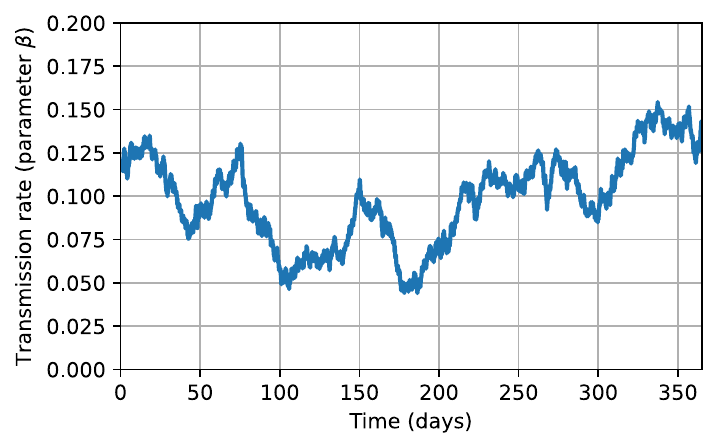}
\caption{}\label{fig:case0_beta}
\end{subfigure}
\begin{subfigure}{0.33\linewidth}
\centering
\includegraphics[width=\linewidth,keepaspectratio]{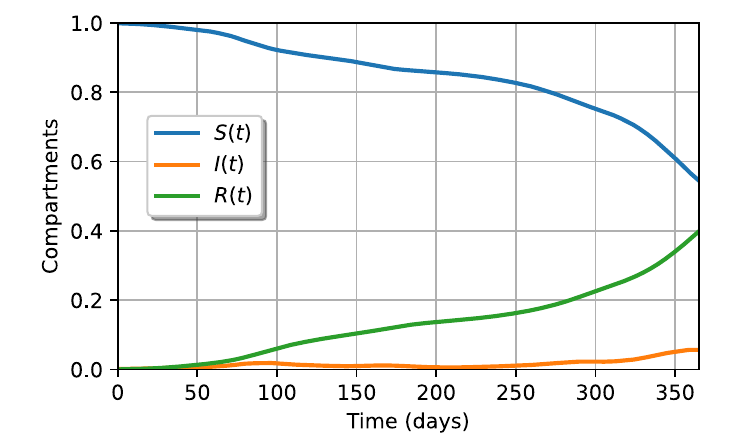}
\caption{}\label{fig:case0_data}
\end{subfigure}
\begin{subfigure}{0.33\linewidth}
\centering
\includegraphics[width=\linewidth,keepaspectratio]{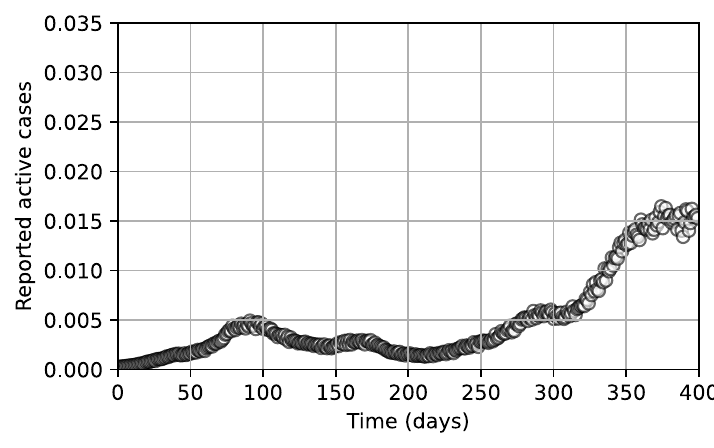}
\caption{}\label{fig:case0_data}
\end{subfigure}
\\
\begin{subfigure}{0.33\linewidth}
\includegraphics[width=\linewidth,keepaspectratio]{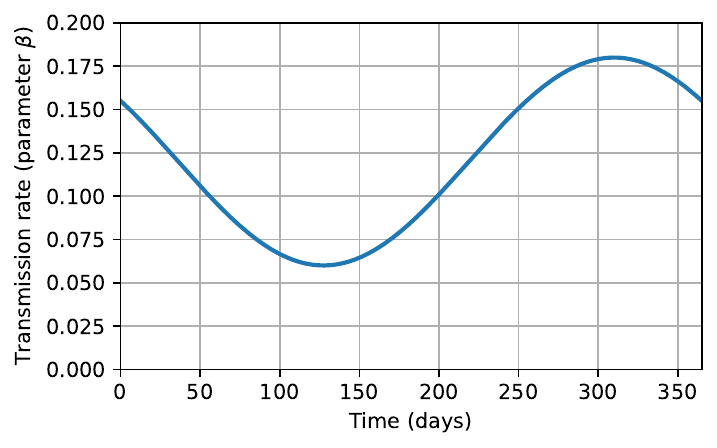}
\caption{}\label{fig:case1_beta}
\end{subfigure}
\begin{subfigure}{0.33\linewidth}
\centering
\includegraphics[width=\linewidth,keepaspectratio]{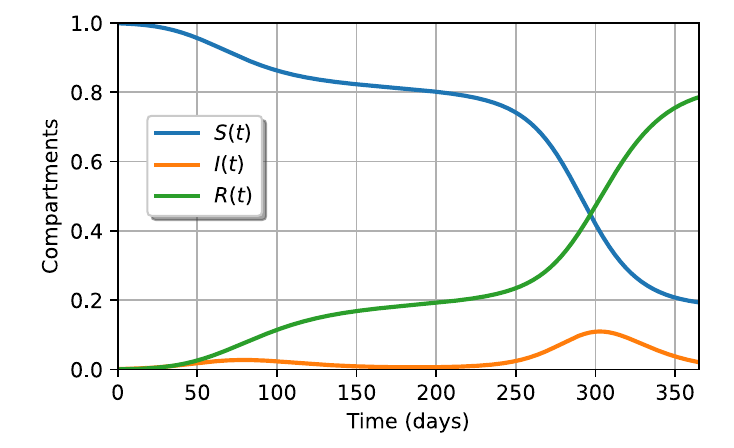}
\caption{}\label{fig:case0_data}
\end{subfigure}
\begin{subfigure}{0.33\linewidth}
\centering
\includegraphics[width=\linewidth,keepaspectratio]{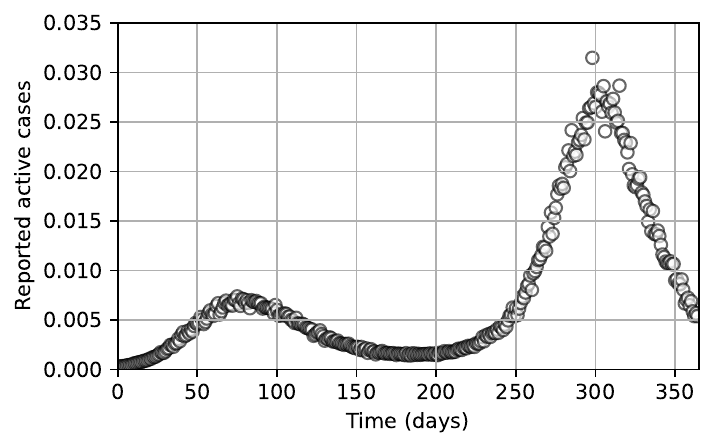}
\caption{}\label{fig:case1_data}
\end{subfigure}
\\
\begin{subfigure}{0.33\linewidth}
\includegraphics[width=\linewidth,keepaspectratio]{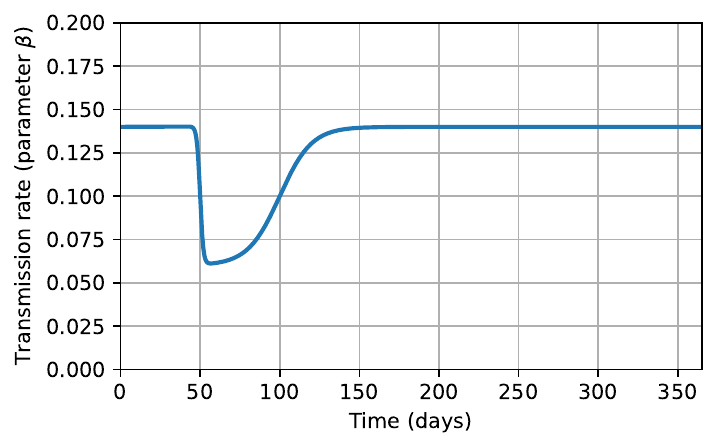}
\caption{}\label{fig:case2_beta}
\end{subfigure}
\begin{subfigure}{0.33\linewidth}
\centering
\includegraphics[width=\linewidth,keepaspectratio]{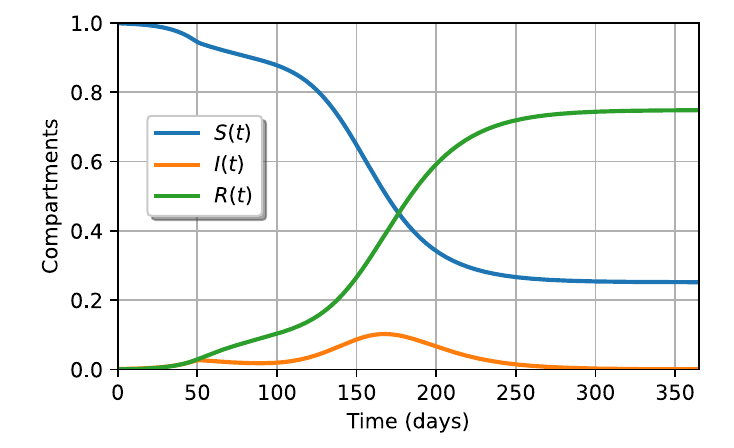}
\caption{}\label{fig:case0_data}
\end{subfigure}
\begin{subfigure}{0.33\linewidth}
\centering
\includegraphics[width=\linewidth,keepaspectratio]{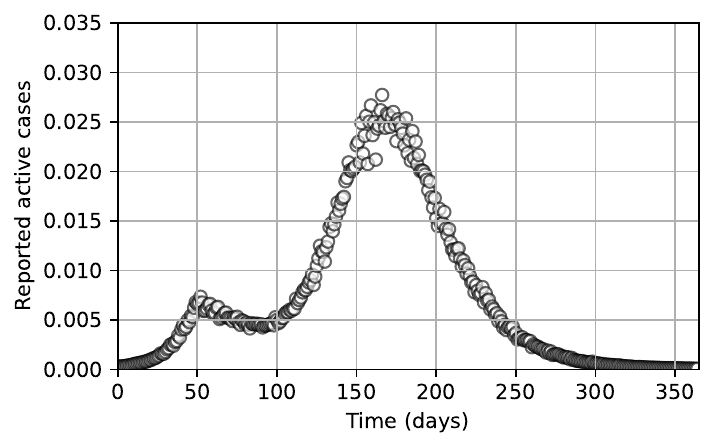}
\caption{}\label{fig:case2_data}
\end{subfigure}
\caption{The prescribed functions of the infection rate $\beta(t)$ (left column), the compartments $S(t)$, $I(t)$, and $R(t)$ (middle column) and the synthetic data (right column). The first row, panels (a), (b), and (c) show synthetic data case 1 with a randomly varying infection rate. The second row, panels (d), (e), and (f) show synthetic data case 2 with a seasonally varying infection rate. The third row in panels (g), (h), and (i) show synthetic data case 3 with a simulated abrupt lockdown with the subsequent gradual removal of restrictions. All synthetic data represents detected active cases, which here accounts for 25\% of the infectious compartment, sampled once per day, and corrupted by 5\% multiplicative noise.}
\end{figure}

Here we demonstrate the capability of the computational framework in jointly estimating the above time-invariant parameters, system states and time-varying parameter. The three cases considered here seek to capture multiple waves of infection, which a deterministic SIR model with time-invariant parameters is unable to adequately model. In addition to representing three unique characteristics of disease transmission that were prevalent during the COVID-19 pandemic, these distinct forms of a time-varying infection rate are formulated to test different abilities of the Bayesian computational framework.  The first case in Figure \ref{fig:case0_beta}, we see an infection rate whose evolution in time is generated from a Wiener process. The random fluctuations in the infection rate at each individual timestep are relatively small but compounded over time leading to significant variability over the course of the simulated one-year period. The resulting data generated using this infection rate parameter, shown in Figure \ref{fig:case0_data}, is by comparison remains relatively smooth over time. This will test the ability of the algorithm to track the highly oscillatory nature of the unobserved time-varying parameter through indirect partial observations of the infectious compartment. The two subsequent cases, the infection rate parameter in Figures \ref{fig:case1_beta} and \ref{fig:case2_beta} vary according to prescribed continuous functions in time. Through partial observations in Figure \ref{fig:case1_data}, the algorithm will be tasked with tracking a large, but gradual variation in the value of the time-varying parameter value. Similarly, from the observations in Figure \ref{fig:case2_data} a similar change in parameter value will have to be tracked but the simulated lockdown results in a much more sudden change. The ability to track the time-varying changes in the infection rate parameter value hinges critically on the estimation of the optimal strength of the artificial noise, enabled by the Bayesian computational framework.

\subsubsection{Synthetic data case 1: random variations}\label{synthetic_random}
The case of a noisy infection rate illustrated in Figures \ref{fig:case0_beta} and \ref{fig:case0_data} is presented first. Given that the fluctuations in the time-varying infection rate being given by a Wiener process, in this case alone, the data-generating model is identical to the model being used for inference in Eq. (\ref{eq:sirb}). Hence, in this case, there is an underlying known true value for each parameter. In the subsequent cases, where the infection rate is defined by a time-varying function, there will be no such ground truth for the artificial noise strength, $q_\xi$. 

\begin{figure}[H]
\centering
\includegraphics[width=0.8\textwidth]{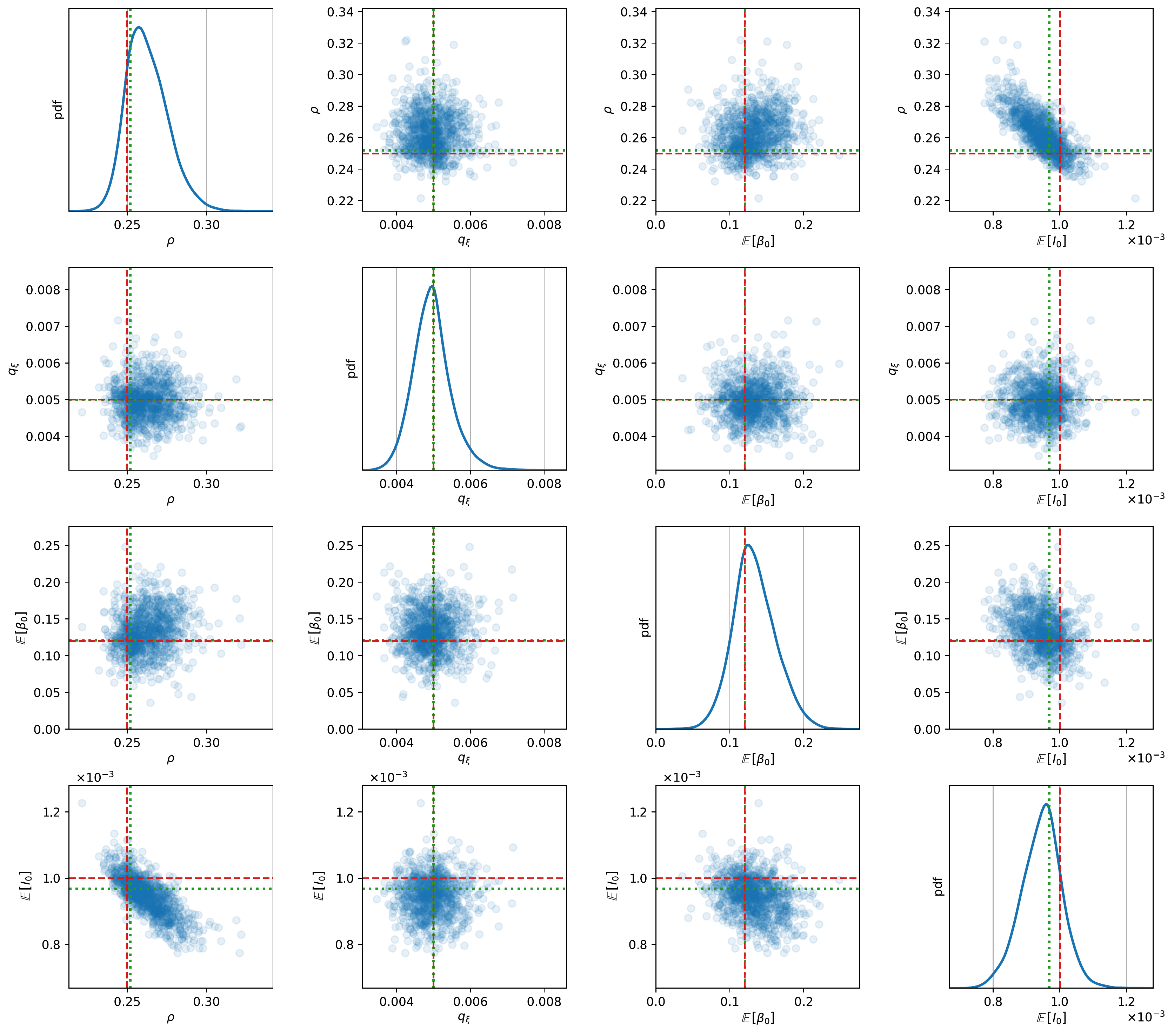}
\caption{Marginal parameter posterior pdfs and scatterplots of pairwise joint parameter posterior pdfs. True parameter values are indicated by dashed lines ($\textcolor{tabred}{\blacksquare}$) and MAP estimates are indicated by dotted lines ($\textcolor{tabgreen}{\blacksquare}$).}
\label{fig:case00_pdf}
\end{figure}

Figure \ref{fig:case00_pdf}, which presents a matrix of the marginal and pairwise joint posterior pdfs of the time-invariant parameters. The MAP estimates are indicated by a dotted line, which can be visually compared against the true parameter value, indicated by a dashed line. All parameter values are captured within the support of the posterior pdfs, with the MAP estimate being closely aligned with the true parameter value. Furthermore, note that the posterior pdfs appear to be nearly Gaussian, with the ratio of detected-to-undetected cases parameter, $\rho$ exhibiting high correlation with the estimate of the initial mean number of infectious individuals.

All true parameter values are captured within the support of the posterior pdfs, and the MAP values all appear to be all closely aligned with the true parameter values. To provide a numerical comparison against the ground truth value of the parameters, the MAP estimate, the mean, and standard deviation are summarized in Table \ref{table:0}.

\begin{table}[!h]
\centering
\begin{tabular}{ p{4cm} | p{2cm} p{2cm} p{2cm} p{2cm}}
\hline
Parameter \Big.  & $\rho$ & $q_\xi$  & $\mathbb{E}[\beta_0]$ &  $\mathbb{E}[I_0]$ \\	
\hline	
 Ground truth & 0.2500 & 5.000  {\footnotesize $\times 10^{-3}$} & 0.1200 & 1.000  {\footnotesize $\times 10^{-3}$}\\
 MAP &  0.2519 & 4.995 {\footnotesize $\times 10^{-3}$} &	0.1208 &0.968 {\footnotesize $\times 10^{-3}$} \\	
Mean & 0.2627  & 4.957  {\footnotesize $\times 10^{-3}$} &0.1336 & 0.945 {\footnotesize $\times 10^{-3}$} \\
 Standard deviation & 0.0130 &  0.516 {\footnotesize $\times 10^{-3}$} & 0.0289 & 0.060 {\footnotesize $\times 10^{-3}$} \\
\hline		
\hline		
\end{tabular}
\caption{Pertinent statistics of the parameter posterior pdfs of time-invariant parameters in Figure \ref{fig:case00_pdf}.}
\label{table:0}
\end{table}

\begin{figure}[H]
\centering
\begin{subfigure}{\linewidth}
\centering
\includegraphics[width=0.8\linewidth,keepaspectratio]{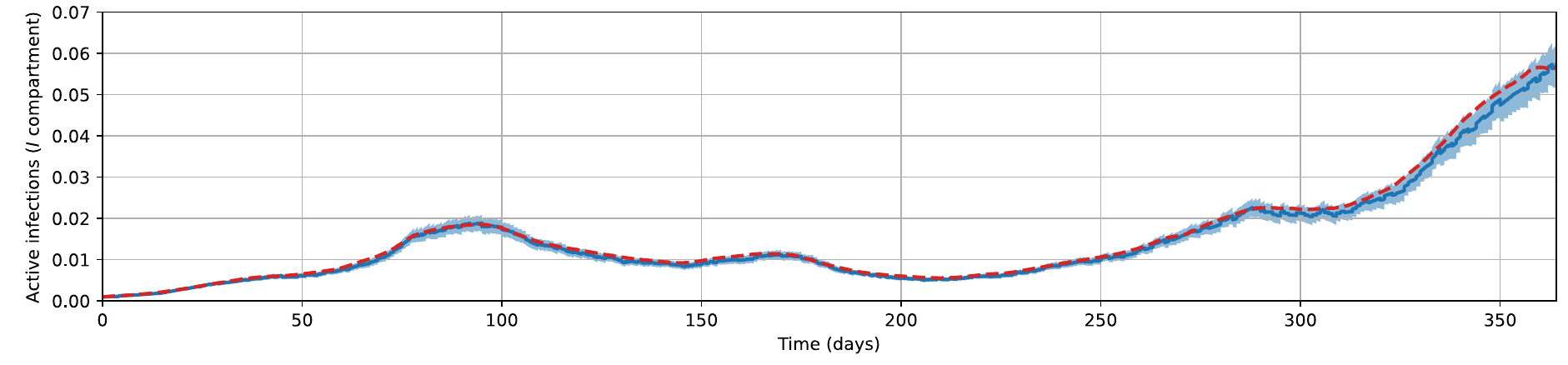}
\caption{}\label{fig:case00_i}
\end{subfigure}
\\
\begin{subfigure}{\linewidth}
\centering
\includegraphics[width=0.8\linewidth,keepaspectratio]{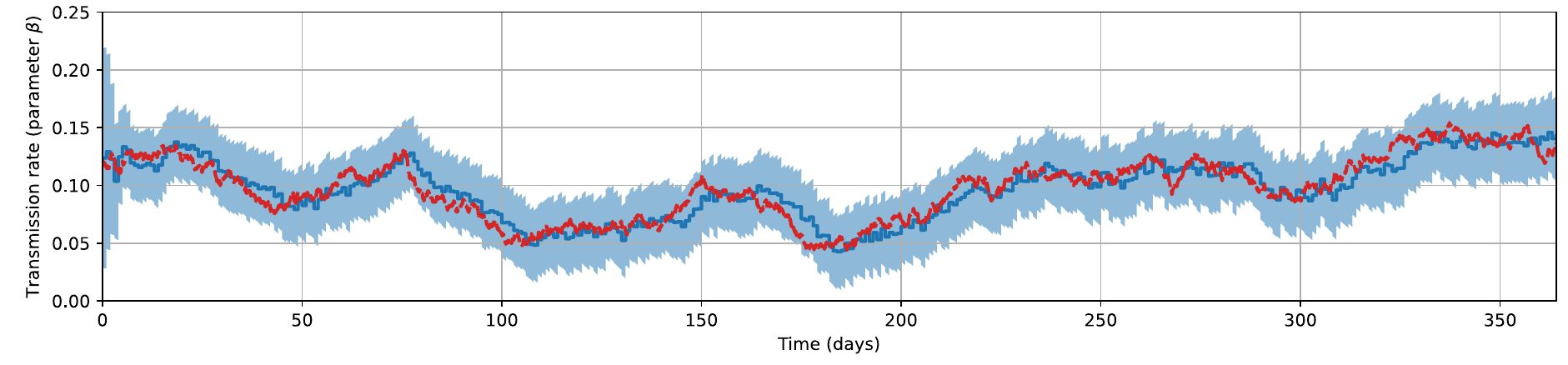}
\caption{}\label{fig:case00_beta}
\end{subfigure}
\\
\begin{subfigure}{\linewidth}
\centering
\includegraphics[width=0.8\linewidth,keepaspectratio]{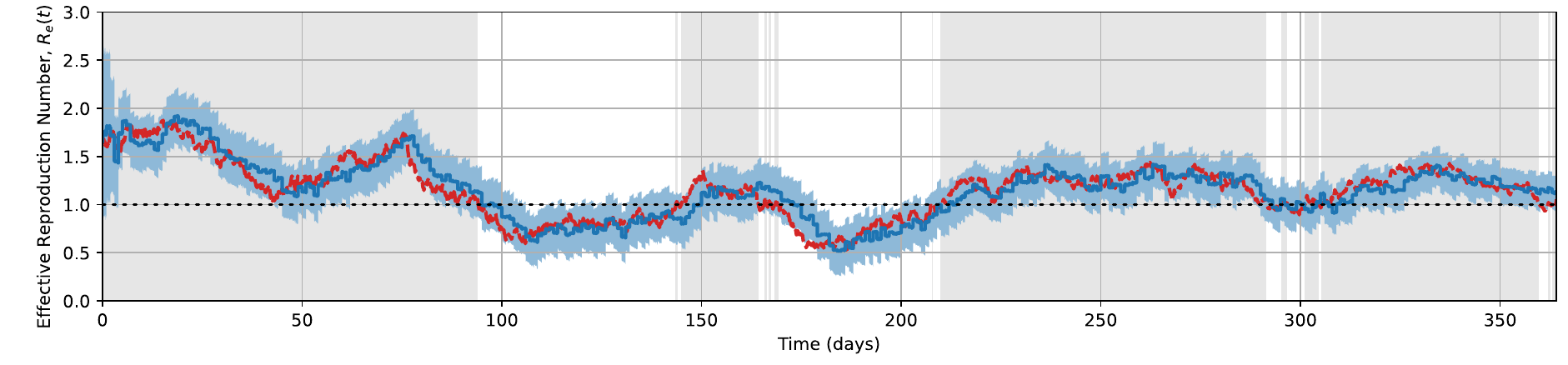}
\caption{}\label{fig:case00_re}
\end{subfigure}
\caption{State estimation results $\text{p}(\mathbf{x}_1,\hdots,\mathbf{x}_K \vert \mathbf{D}, \bm{\Phi}_s^{\text{MAP}},\mathcal{M})$ for (a) the infectious compartment, $I$, (b) the time-varying infection rate parameter, $\beta$, and (c) the effective reproduction number $R_e$. The ground truth is indicated by a dashed line ($\textcolor{tabred}{\blacksquare}$), the mean estimates are indicated by a solid line ($\textcolor{tabblue}{\blacksquare}$), and the shaded area reflects the mean $\pm$ 3 standard deviations.}\label{fig:case00_state}
\end{figure}

In Figure \ref{fig:case00_state}, we provide the mean state estimate and a three standard deviation uncertainty interval for the infectious compartment $I(t)$ (panel \ref{fig:case00_i}) and the time-varying infection rate parameter $\beta(t)$ (panel \ref{fig:case00_beta}), plotted against the known ground truth from the data-generating model. We also plot the effective reproduction number, a quantity that is derived using the joint state estimates of $S(t)$ and $\beta(t)$ (panel \ref{fig:case00_re}). We therefore omit the state estimation results for the susceptible and recovered compartment for brevity. The results shown for the augmented state are conditioned on the maximum a posteriori (MAP) estimate of the time-invariant parameters.

\begin{figure}[!htb]
\centering
\begin{subfigure}{0.33\linewidth}
\centering
\includegraphics[width=\linewidth,keepaspectratio]{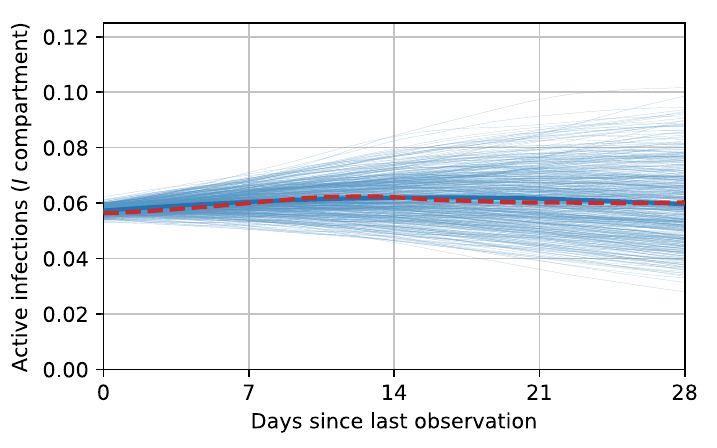}
\caption{}\label{fig:case07_i}
\end{subfigure}
\begin{subfigure}{0.33\linewidth}
\centering
\includegraphics[width=\linewidth,keepaspectratio]{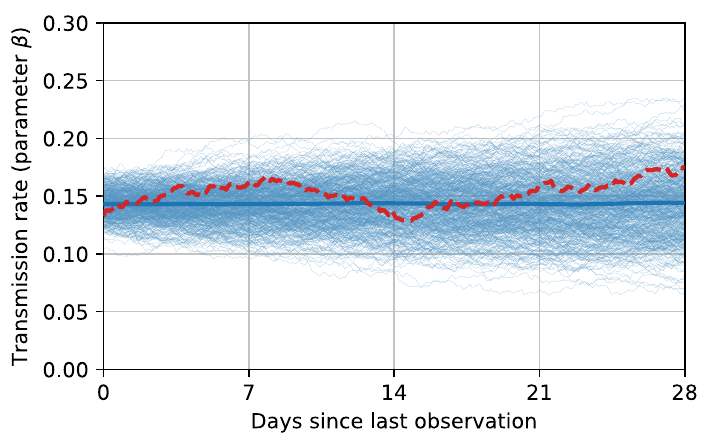}
\caption{}\label{fig:case07_beta}
\end{subfigure}
\caption{Four week forecast after assimilating the last available data point on day 365. (a) the infectious compartment, $I$ and (b) the time-varying infection rate parameter, $\beta$ illustrated using 500 Monte Carlo samples. The ground truth is indicated by a dashed line ($\textcolor{tabred}{\blacksquare}$), the mean estimates are indicated by a solid line ($\textcolor{tabblue}{\blacksquare}$).}\label{fig:forecast}
\end{figure}

Critically, the joint estimation of the time-invariant parameters, the time-varying parameters and the system states permits the forecasting of future cases under uncertainty. Figure \ref{fig:forecast} illustrates a four-week forecasts of the $I$ compartment, representing the true number of active cases in panel \ref{fig:case07_i}, and the time-varying infection rate in panel \ref{fig:case07_beta}. 
The forecast is obtained by integrating the full nonlinear SIR model.
 The resulting uncertainty bounds grow in time, such that they envelope the possible realizations of the randomly varying infection rate.  While the agreement of the predictions shown in panel \ref{fig:case07_i}, hinges on the successful estimation of the $I$ compartment leading up to the data cut-off, it is also contingent upon the specific realization of $\beta$. In this case, the future values of $\beta$ happened to be such that the mean estimate exhibits reasonable agreement. The agreement of the mean estimate of $I$ will not be as close for all realizations of $\beta$. However, the uncertainty bounds grow in time, enveloping the possible trajectories of infectious compartment $I$.

Continuing with this synthetic data case, we intend to highlight the critical importance of the framework's ability to estimate all the model parameters. Through a series of examples, we illustrate the negative impact that imposing incorrect assumptions regarding parameter values has on the joint estimates of the system state, time-varying parameter, and time-invariant parameters. This will be shown through the inability to accurately capture the ground truth in the presence of such assumptions. 

\begin{figure}[H]
\centering
\begin{subfigure}{0.33\linewidth}
\includegraphics[width=\linewidth,keepaspectratio]{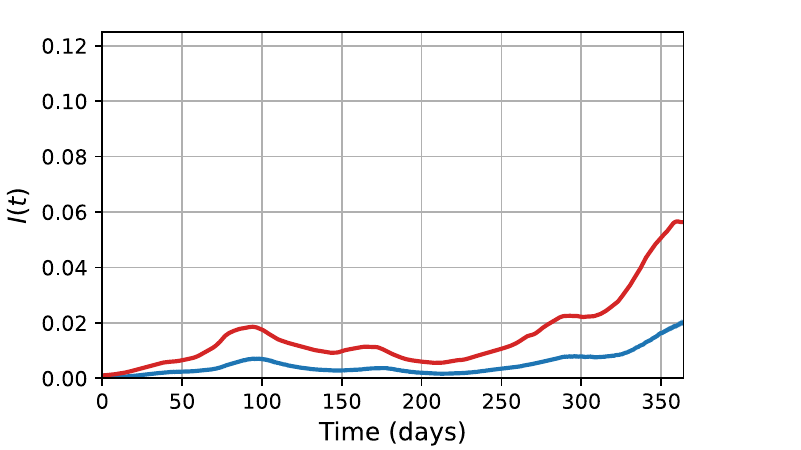}
\caption{}\label{fig:app_beta3}
\end{subfigure}
\quad
\begin{subfigure}{0.33\linewidth}
\centering
\includegraphics[width=\linewidth,keepaspectratio]{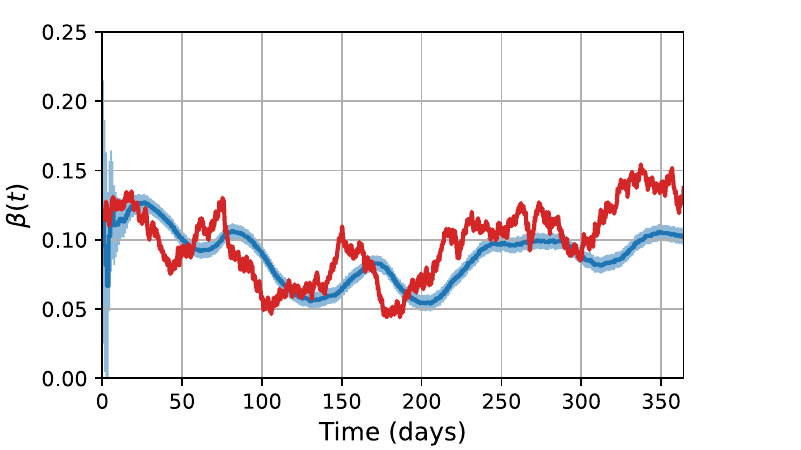}
\caption{}\label{fig:app_beta4}
\end{subfigure}
\\
\begin{subfigure}{0.33\linewidth}
\includegraphics[width=\linewidth,keepaspectratio]{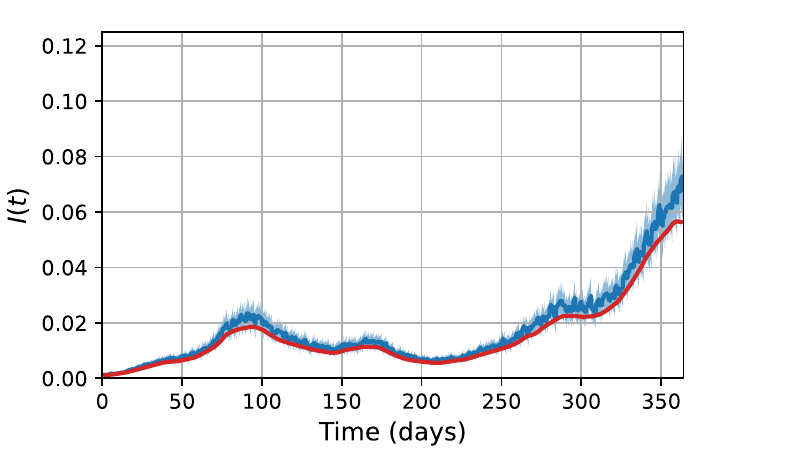}
\caption{}\label{fig:app_beta1}
\end{subfigure}
\quad
\begin{subfigure}{0.33\linewidth}
\centering
\includegraphics[width=\linewidth,keepaspectratio]{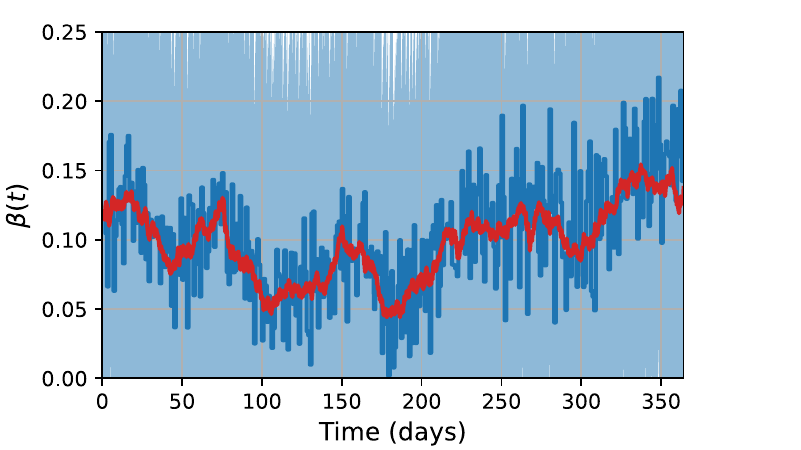}
\caption{}\label{fig:app_beta2}
\end{subfigure}
\caption{State estimation results $\text{p}(\mathbf{x}_1,\hdots,\mathbf{x}_K \vert \mathbf{D}, \bm{\Phi}_s^{\text{MAP}},\mathcal{M})$ for the infectious compartment, $I$ (left column) and the time-varying infection rate parameter, $\beta$ (right column). The ground truth is indicated by a dashed line ($\textcolor{tabred}{\blacksquare}$), the mean estimates are indicated by a solid line ($\textcolor{tabblue}{\blacksquare}$), and the shaded area reflects the mean $\pm$ 3 standard deviations. Recall the generating value of the artificial noise strength, $q_\xi$ is 0.005. Panels (a) and (b) have $q_\xi = 0.001$. Panels (c) and (d) have $q_\xi = 0.01$.}\label{fig:app1}
\end{figure}

The time-varying infection rate parameter $\beta(t)$ was generated from a Wiener process having a noise strength with a magnitude of 0.005. Here we consider the results in the case where rather than estimating this critical parameter through the inference procedure, we relied on manual tuning, selecting sub-optimal values. We illustrate how a low value of 0.001 and a high value of 0.01 affect the results in Figure \ref{fig:app1}. For this illustrative example, we have taken for granted that we know the initial conditions, thus the only time-invariant parameter that remains to be estimated is the ratio of detected to undetected infectious cases, $\rho$. Panels \ref{fig:app_beta3} and \ref{fig:app_beta4} depict the consequences of insufficient perturbation strength for time-varying parameter estimation. The  the state estimates of $\beta$ clearly illustrate how the low artificial noise strength severely limits the ability of the filter to track the true value of $\beta$, with the estimate of $\beta$ being much more gradually evolving than the truth (panel \ref{fig:app_beta4}). Through the joint estimation of the ratio of detected to undetected infectious cases, the model still manages to exhibit reasonable data-fit. However, the resulting effect leads to a severe underestimate of the infectious compartment (panel \ref{fig:app_beta3}). The case where the artificial noise strength is too high is illustrated in panels \ref{fig:app_beta1} and \ref{fig:app_beta2}. Here, the true infectious compartment is tracked well by the state estimation procedure (though being noisy). This noise is even more apparent in the estimates of the infection rate parameter (panel \ref{fig:app_beta2}), which tracks the trend well. However, there is significant noise in the mean estimate, and the plotted three standard deviation uncertainty intervals indicate significant uncertainty in the estimate.

%

\begin{figure}[H]
\centering
\begin{subfigure}{0.33\linewidth}
\includegraphics[width=\linewidth,keepaspectratio]{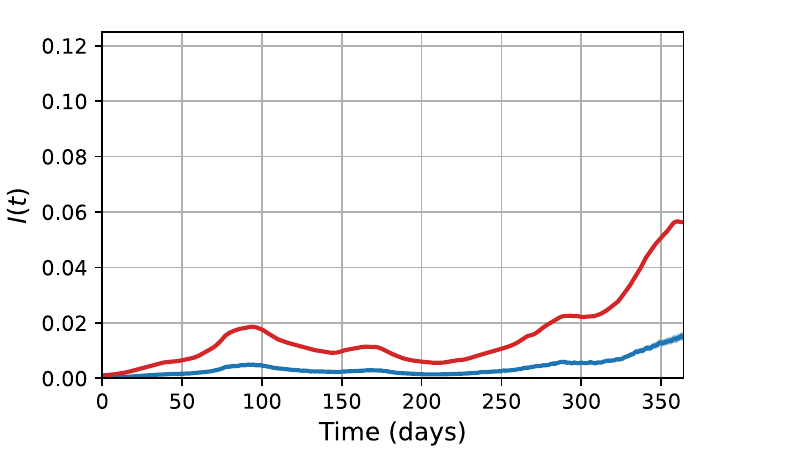}
\caption{}\label{fig:app_beta13}
\end{subfigure}
\quad
\begin{subfigure}{0.33\linewidth}
\centering
\includegraphics[width=\linewidth,keepaspectratio]{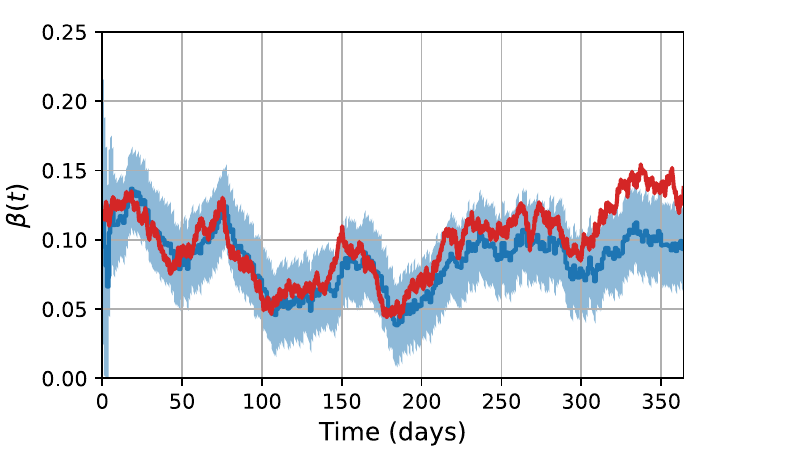}
\caption{}\label{fig:app_beta14}
\end{subfigure}
\\
\begin{subfigure}{0.33\linewidth}
\includegraphics[width=\linewidth,keepaspectratio]{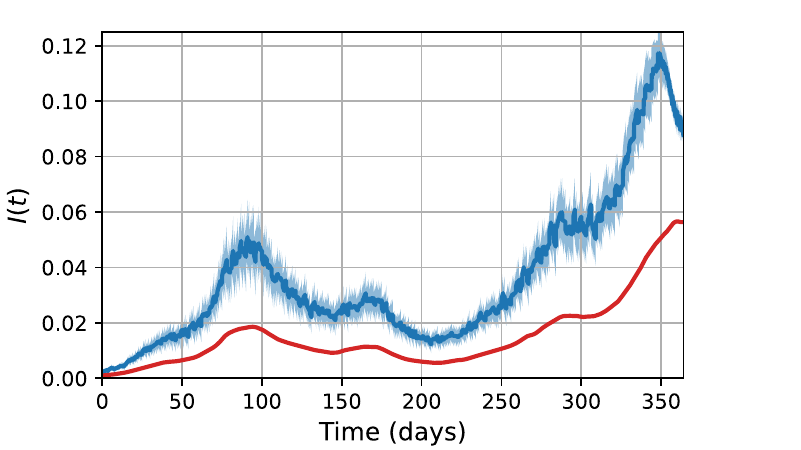}
\caption{}\label{fig:app_beta15}
\end{subfigure}
\quad
\begin{subfigure}{0.33\linewidth}
\centering
\includegraphics[width=\linewidth,keepaspectratio]{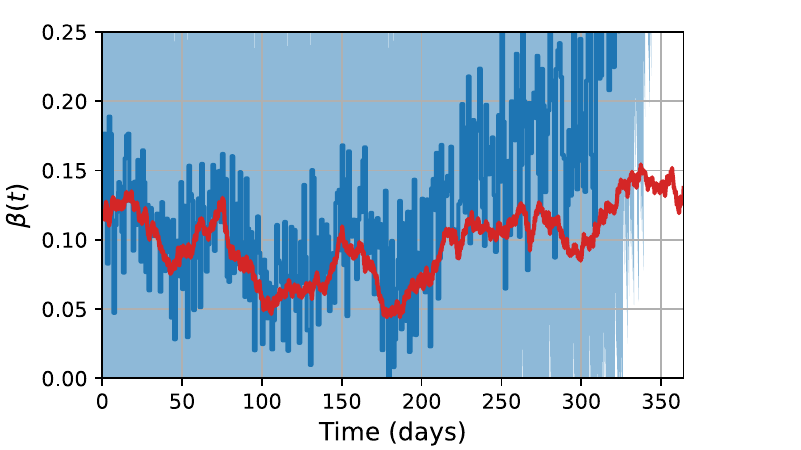}
\caption{}\label{fig:app_beta16}
\end{subfigure}
\caption{State estimation results $\text{p}(\mathbf{x}_1,\hdots,\mathbf{x}_K \vert \mathbf{D}, \bm{\Phi}_s^{\text{MAP}},\mathcal{M})$ for the infectious compartment, $I$ (left column) and the time-varying infection rate parameter, $\beta$ (right column). The ground truth is indicated by a dashed line ($\textcolor{tabred}{\blacksquare}$), the mean estimates are indicated by a solid line ($\textcolor{tabblue}{\blacksquare}$), and the shaded area reflects the mean $\pm$ 3 standard deviations. Recall the generating value of the ratio of detected to undetected cases, $\rho$ is 0.25. Panels (a) and (b) have $\rho = 1.0$. Panels (c) and (d) have $\rho = 0.1$.}\label{fig:app2}
\end{figure}

To complement this exercise, we now propose a model where we assume we know $\rho$ (the ratio of detected to undetected cases), but decide to estimate the artificial noise strength needed to track the evolution of the infection rate, $\beta$ in time in Figure \ref{fig:app2}. Recall, the data was generated using a 25\% detection rate. Hence, we will see how the estimation results change if we were to assume a 100\% detection rate, or 10\% detection rate. Intuitively, we will expect the assumed 100\% detection rate will result in an underestimate of the true number of infections, and conversely, the 10\% detection rate will result in an overestimate. This conjecture is confirmed in panels \ref{fig:app_beta13} and \ref{fig:app_beta15}. What is of interest here, is understanding how these assumptions influence the estimation of the infection rate parameter, from which the effective reproduction number is derived. In panel \ref{fig:app_beta14}, we can see that overestimating the detection ratio has a minimal influence on the ability to track the infection rate, whereas in panel \ref{fig:app_beta16}, underestimating the detection ratio results in poor tracking. This occurs because of the estimation of the artificial noise strength. Recall, the generating value of the artificial noise strength $q_\xi$ was $5.000 \times 10^{-3}$. The results in panel \ref{fig:app_beta14}, is obtained from a MAP estimate of $4.351\times 10^{-3}$, whereas the MAP estimate corresponding to the results in panel \ref{fig:app_beta16} was $6.585\times 10^{-2}$ , an order of magnitude larger. The reason for the large value is not because the framework was unable to estimate the parameter, but rather because in this situation, a larger magnitude process noise is required in order to achieve satisfactory data-fit. 

\begin{figure}[H]
\centering
\begin{subfigure}{0.33\linewidth}
\includegraphics[width=\linewidth,keepaspectratio]{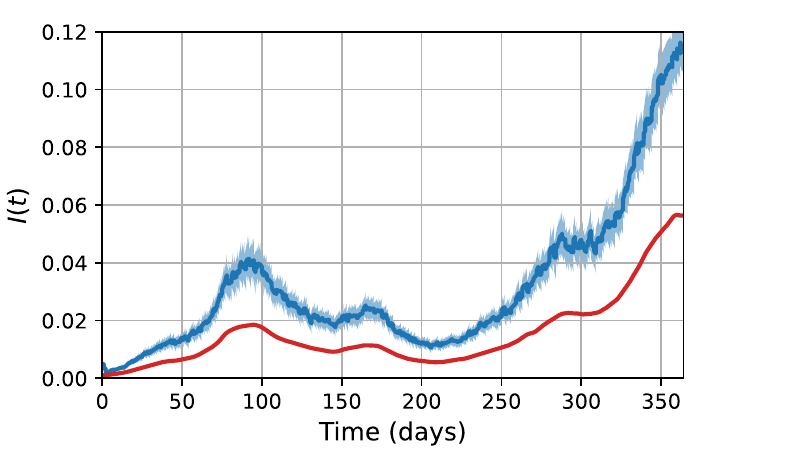}
\caption{}\label{fig:app_beta5}
\end{subfigure}
\quad
\begin{subfigure}{0.33\linewidth}
\centering
\includegraphics[width=\linewidth,keepaspectratio]{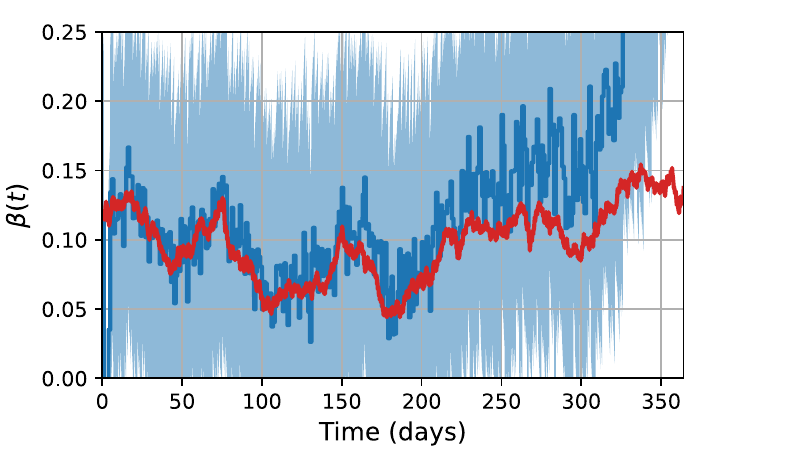}
\caption{}\label{fig:app_beta6}
\end{subfigure}
\\
\begin{subfigure}{0.33\linewidth}
\includegraphics[width=\linewidth,keepaspectratio]{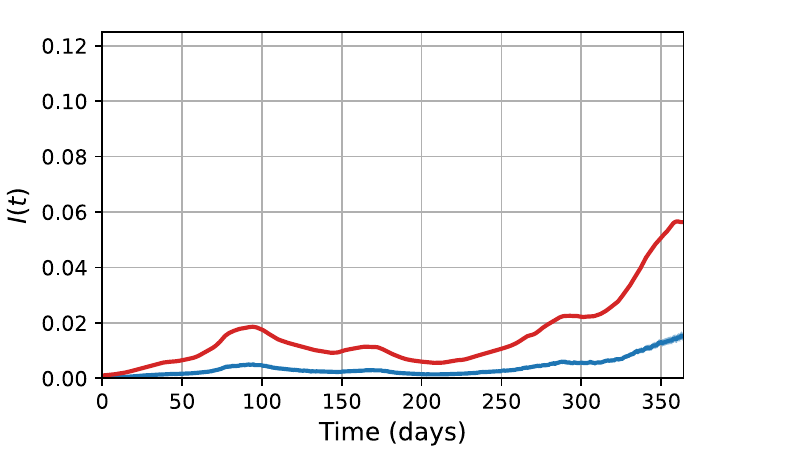}
\caption{}\label{fig:app_beta7}
\end{subfigure}
\quad
\begin{subfigure}{0.33\linewidth}
\centering
\includegraphics[width=\linewidth,keepaspectratio]{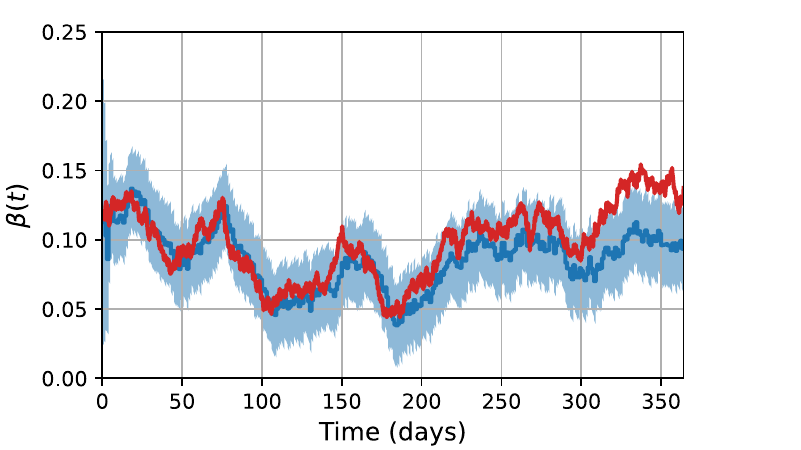}
\caption{}\label{fig:app_beta8}
\end{subfigure}
\\
\begin{subfigure}{0.33\linewidth}
\includegraphics[width=\linewidth,keepaspectratio]{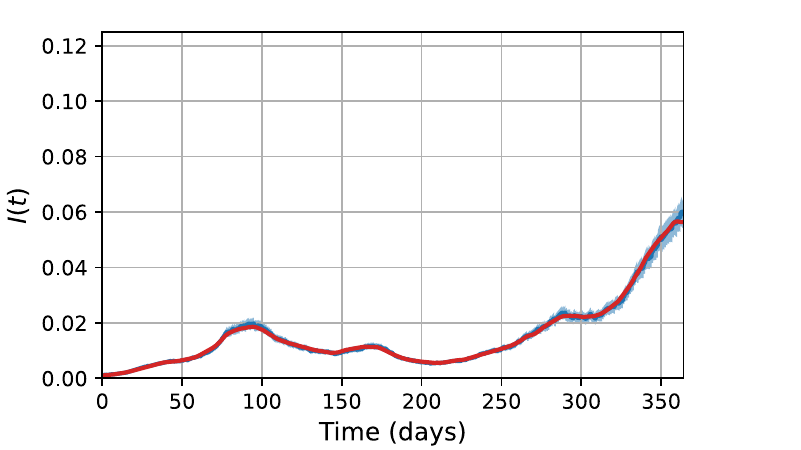}
\caption{}\label{fig:app_beta9}
\end{subfigure}
\quad
\begin{subfigure}{0.33\linewidth}
\centering
\includegraphics[width=\linewidth,keepaspectratio]{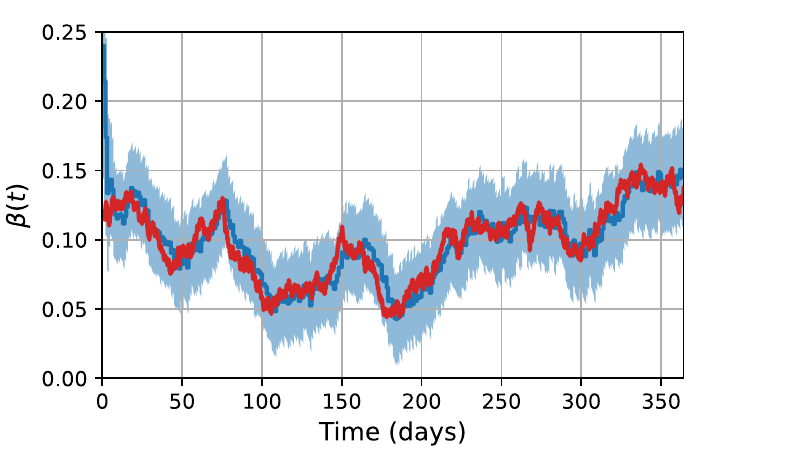}
\caption{}\label{fig:app_beta10}
\end{subfigure}
\\
\begin{subfigure}{0.33\linewidth}
\includegraphics[width=\linewidth,keepaspectratio]{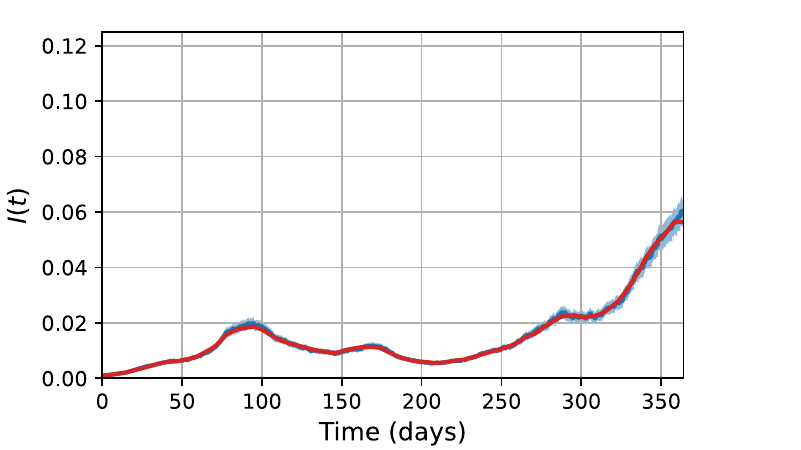}
\caption{}\label{fig:app_beta11}
\end{subfigure}
\quad
\begin{subfigure}{0.33\linewidth}
\centering
\includegraphics[width=\linewidth,keepaspectratio]{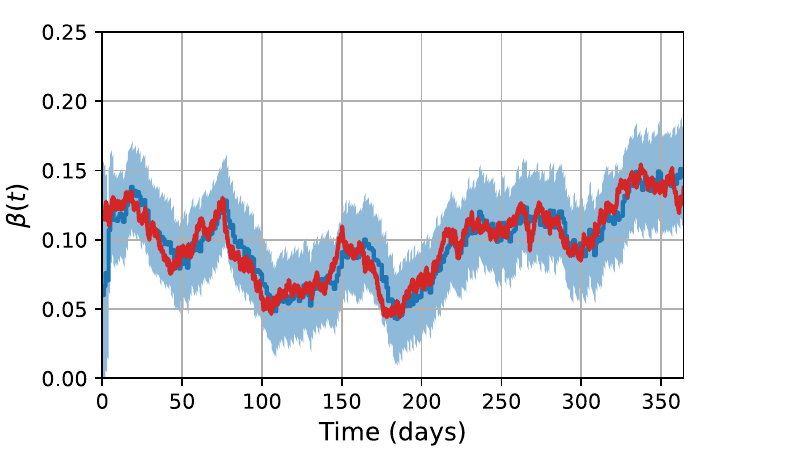}
\caption{}\label{fig:app_beta12}
\end{subfigure}
\caption{State estimation results $\text{p}(\mathbf{x}_1,\hdots,\mathbf{x}_K \vert \mathbf{D}, \bm{\Phi}_s^{\text{MAP}},\mathcal{M})$ for the infectious compartment, $I$ (left column) and the time-varying infection rate parameter, $\beta$ (right column). The ground truth is indicated by a dashed line ($\textcolor{tabred}{\blacksquare}$), the mean estimates are indicated by a solid line ($\textcolor{tabblue}{\blacksquare}$), and the shaded area reflects the mean $\pm$ 3 standard deviations. Recall the initial conditions of the generating model are $I_0 = 0.001$ and $\beta_0 = 0.12$. Panels (a) and (b) have $\mathbb{E}[I_0] = 0.005$ and $\mathbb{E}[\beta_0] = 0.12$. Panels (c) and (d) have $\mathbb{E}[I_0] = 0.0002$ and $\mathbb{E}[\beta_0] = 0.12$. Panels (e) and (f) have $\mathbb{E}[I_0] = 0.001$ and $\mathbb{E}[\beta_0] = 0.24$. Panels (g) and (h) have $\mathbb{E}[I_0] = 0.001$ and $\mathbb{E}[\beta_0] = 0.06$.}\label{fig:app3}
\end{figure}

Finally, we consider a series of combinations of incorrect initial conditions for the augmented state vector in Figure \ref{fig:app3}. In panels \ref{fig:app_beta5} and \ref{fig:app_beta5} we have the correct initial value of $\beta_0 = 0.12$, but an overestimate of $I_0 = 0.01$. In panels \ref{fig:app_beta7} and \ref{fig:app_beta8}, once again, we have the correct initial value of $\beta_0 =0.12$, but an underestimate of $I_0=0.0001$. These four panels exhibit similar behaviour to what was depicted in Figure \ref{fig:app2}. The incorrect initial estimates of the infectious compartment adversely impacts the joint estimates of the state $I(t)$ and the ratio of observed to unobserved cases, $\rho$. Conversely, for a correct initial values of  $I_0 = 0.01$, for an overestimated initial value of $\beta_0 = 0.24$ in panels \ref{fig:app_beta9} and \ref{fig:app_beta10}, or an underestimated initial value of $\beta_0 = 0.06$ in panels \ref{fig:app_beta11} and \ref{fig:app_beta12}, the time-varying parameter is capable of quickly correcting for an erroneous initial condition.

\subsubsection{Synthetic data case 2: seasonal variations}\label{synthetic_seasonal}
The infection rate exhibiting seasonal variation from Figures \ref{fig:case1_beta} and \ref{fig:case1_data} is now considered. Because the infection rate parameter is now defined by a sinusoidal function, there is no longer an underlying true value of the artificial noise strength, $q_\xi$, perturbing $\beta$. Instead, the value of $q_\xi$is estimated such that it introduces minimal artificial dynamics, while being sufficiently large to permit the nonlinear filters to track the value of $\beta$ as it varies in time. This is implicitly imposed by the computation of the likelihood using (see Eq. (\ref{eq:int_ekf})). 

\begin{figure}[H]
\centering
\includegraphics[width=0.8\textwidth]{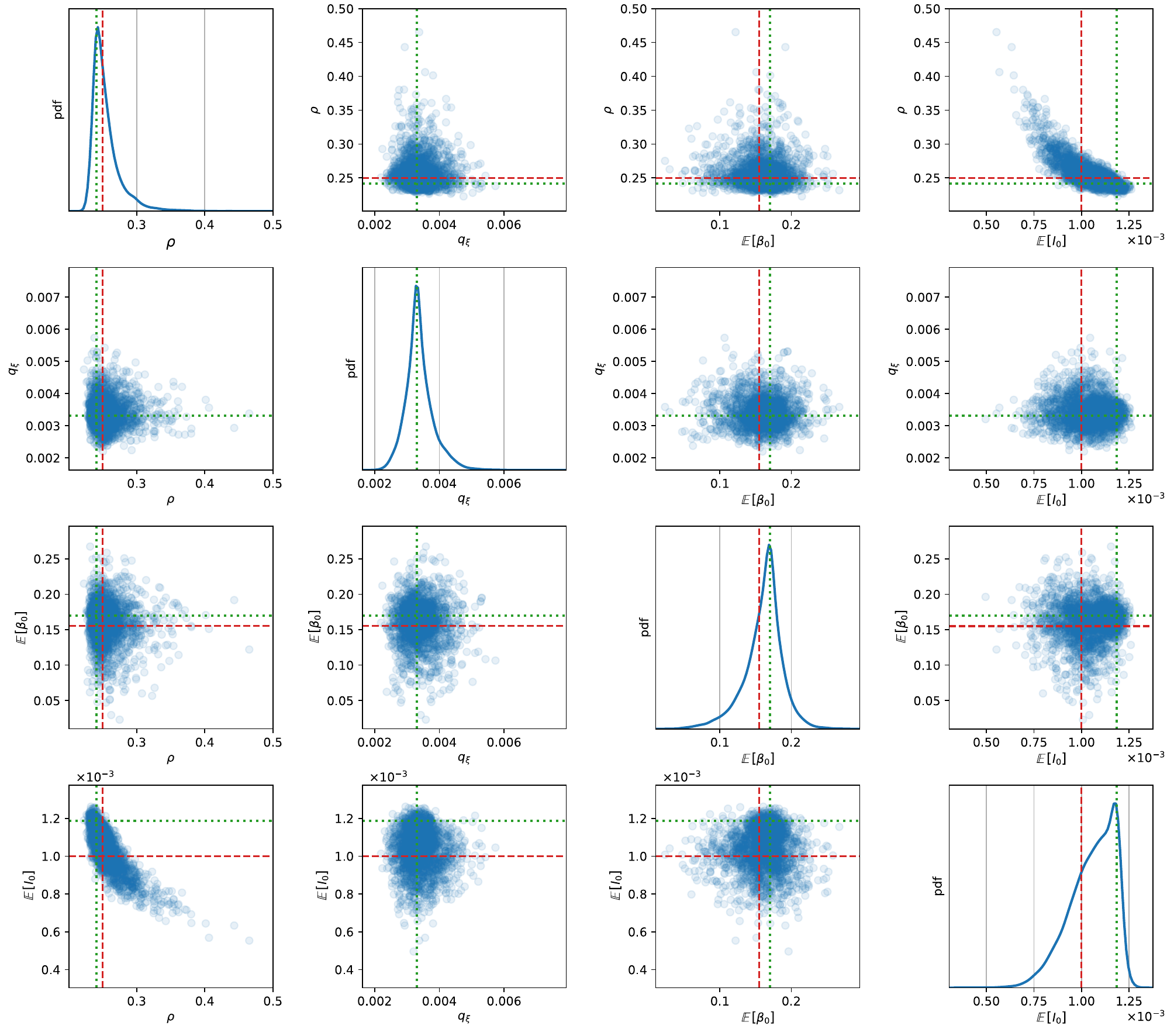}
\caption{Marginal parameter posterior pdfs and scatterplots of pairwise joint parameter posterior pdfs. True parameter values are indicated by dashed lines ($\textcolor{tabred}{\blacksquare}$) and MAP estimates are indicated by dotted lines ($\textcolor{tabgreen}{\blacksquare}$).}
\label{fig:case01_pdf}
\end{figure}

In contrast to the previous case, the posterior pdfs of the time-invariant parameters shown in Figure \ref{fig:case01_pdf} exhibit clear non-Gaussian behaviour. The pdfs of the ratio of detected cases and the initial number of infectious individuals exhibit evident skewness, and increased uncertainty compared to their counterparts in Figure \ref{fig:case00_pdf} for the previous case. Nevertheless, the true parameter values are all captured by the posterior pdfs. Note again that there is no ground truth for the artificial noise strength $q_\xi$.
The mean estimates in Table \ref{table:1} exhibit slightly better agreement with the ground truth values than the MAP estimates plotted in Figure \ref{fig:case01_pdf}. 

\begin{table}[!h]
\centering
\begin{tabular}{ p{4cm} | p{2cm} p{2cm} p{2cm} p{2cm}}
\hline
Parameter \Big.  & $\rho$ & $q_\xi$  & $\mathbb{E}[\beta_0]$ &  $\mathbb{E}[I_0]$ \\	
\hline	
 Ground truth & 0.2500 & - & 0.1551 & 1.000 {\footnotesize $\times 10^{-3}$}\\
 MAP &  0.2414 & 3.307 {\footnotesize $\times 10^{-3}$} &	0.1701 &1.186 {\footnotesize $\times 10^{-3}$} \\	
Mean & 0.2554  & 3.387  {\footnotesize $\times 10^{-3}$} &0.1610& 1.055 {\footnotesize $\times 10^{-3}$} \\
 Standard deviation & 0.0225 &  0.437 {\footnotesize $\times 10^{-3}$} & 0.0264 & 0.118{\footnotesize $\times 10^{-3}$} \\
\hline		
\hline		
\end{tabular}
\caption{Pertinent statistics of the parameter posterior pdfs of time-invariant parameters in Figure \ref{fig:case01_pdf}.}
\label{table:1}
\end{table}

\begin{figure}[H]
\centering
\begin{subfigure}{\linewidth}
\centering
\includegraphics[width=0.8\linewidth,keepaspectratio]{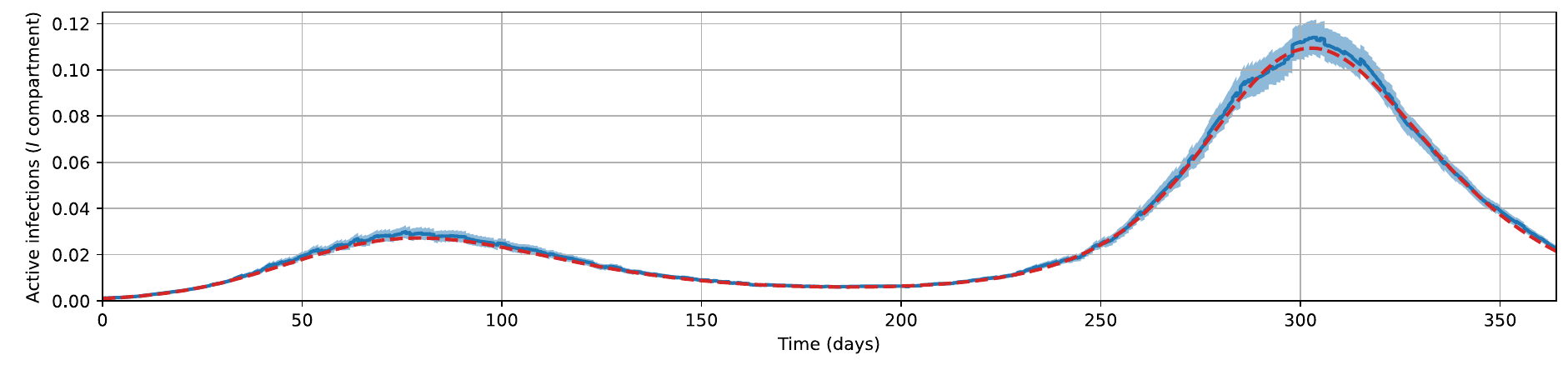}
\caption{}\label{fig:case01_i}
\end{subfigure}
\\
\begin{subfigure}{\linewidth}
\centering
\includegraphics[width=0.8\linewidth,keepaspectratio]{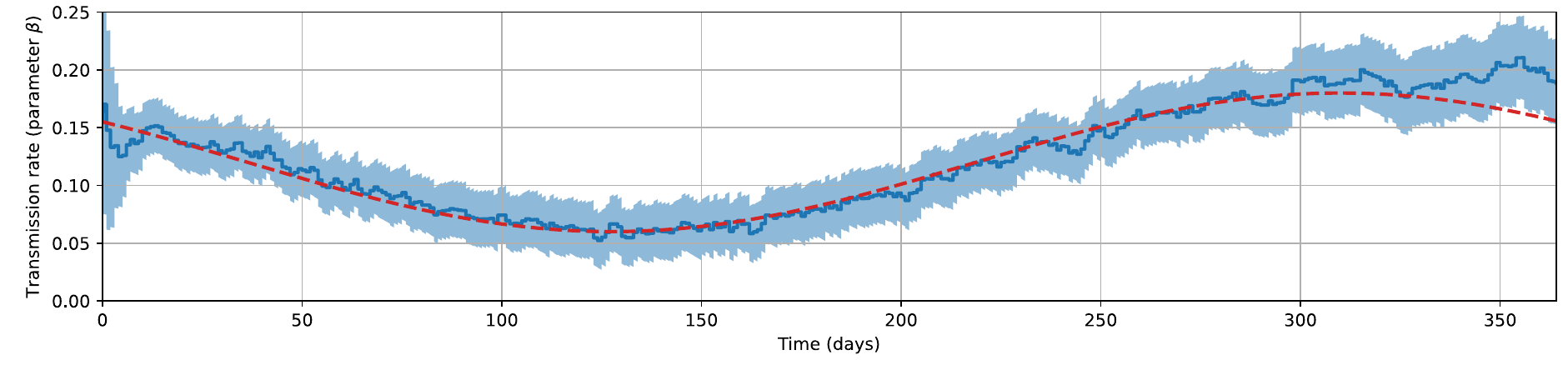}
\caption{}\label{fig:case01_beta}
\end{subfigure}
\\
\begin{subfigure}{\linewidth}
\centering
\includegraphics[width=0.8\linewidth,keepaspectratio]{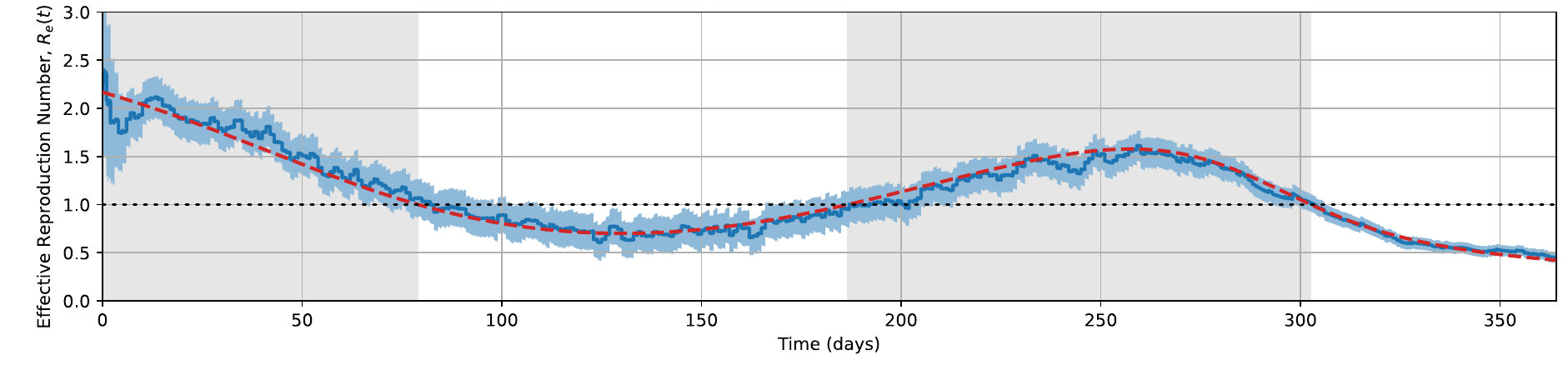}
\caption{}\label{fig:case01_re}
\end{subfigure}
\caption{State estimation results $\text{p}(\mathbf{x}_1,\hdots,\mathbf{x}_K \vert \mathbf{D}, \bm{\Phi}_s^{\text{MAP}},\mathcal{M})$ for (a) the infectious compartment, $I$, (b) the time-varying infection rate parameter, $\beta$, and (c) the effective reproduction number $R_e$. The ground truth is indicated by a dashed line ($\textcolor{tabred}{\blacksquare}$), the mean estimates are indicated by a solid line ($\textcolor{tabblue}{\blacksquare}$), and the shaded area reflects the mean $\pm$ 3 standard deviations.}\label{fig:case01_state}
\end{figure}

In Figure \ref{fig:case01_state}, the resulting state estimation results performed using the MAP estimate of the time-invariant parameter tracks the gradual change in the infection rate reasonably well. While the tracking of $\beta(t)$ (in panel \ref{fig:case01_beta}) appears to degrade near the end of the simulation, the infectious compartment, $I(t)$, and the the effective reproduction number, $R_e(t)$, are estimated accurately (having low uncertainty in their estimates) as shown in panels \ref{fig:case01_i} and \ref{fig:case01_re}, respectively.

\subsubsection{Synthetic data case 3: abrupt lockdown measures}\label{synthetic_lockdown}
\begin{figure}[H]
\centering
\includegraphics[width=0.8\textwidth]{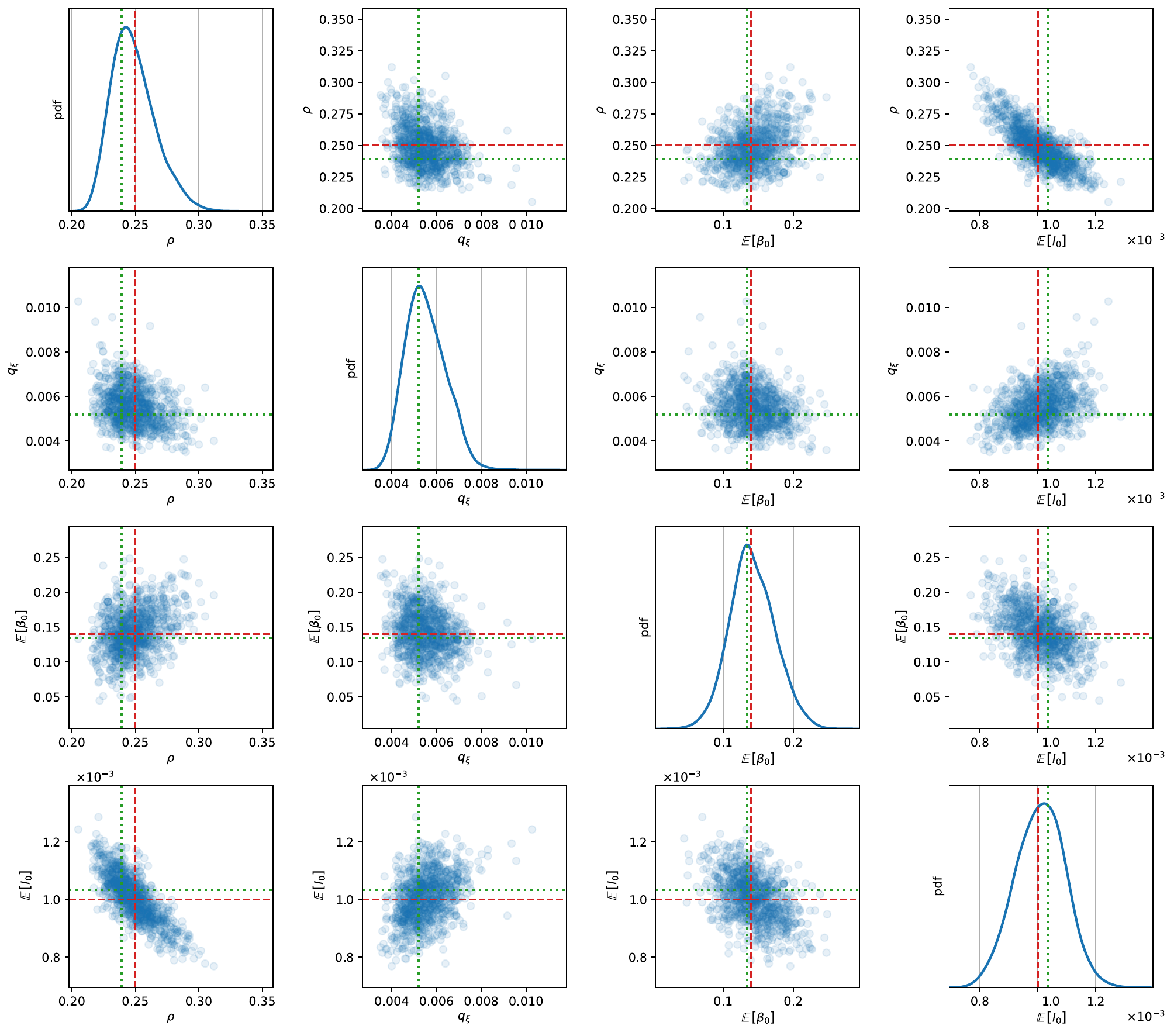}
\caption{Marginal parameter posterior pdfs and scatterplots of pairwise joint parameter posterior pdfs. True parameter values are indicated by dashed lines ($\textcolor{tabred}{\blacksquare}$) and MAP estimates are indicated by dotted lines ($\textcolor{tabgreen}{\blacksquare}$).}
\label{fig:case02_pdf}
\end{figure}

The following data represents a sudden drop in the infection rate parameter caused by introduction of public health measures such as social distancing or lockdown policies. This sudden drop is followed by a gradual return to the baseline parameter value due to the gradual relaxation of the measures. In a similar fashion to the previous example, the infection rate here in the data-generating model is given by an explicit function of time. Due to the significant and sudden change in parameter value, an optimal artificial noise strength value would permit the signal to quickly adjust to the sudden change in the time-varying parameter while minimally inflating the process noise.
The parameter posterior pdfs are presented in Figure \ref{fig:case02_pdf}, and summarized in Table \ref{table:2}. Note again in this case, due to the skewness of the posterior pdf, the mean exhibits slightly closer agreement with the ground truth of the time-invariant parameters than the MAP estimates. As one would expect, the artificial noise strength $q_\xi$ has a larger value in this example, than was reported for the more gradual seasonal variations of the infection rate.


\begin{table}[!h]
\centering
\begin{tabular}{ p{4cm} | p{2cm} p{2cm} p{2cm} p{2cm}}
\hline
Parameter \Big.  & $\rho$ & $q_\xi$  & $\mathbb{E}[\beta_0]$ &  $\mathbb{E}[I_0]$ \\	
\hline	
 Ground truth & 0.2500 & - & 0.1400 & 1.000 {\footnotesize $\times 10^{-3}$}\\
 MAP &  0.2392 & 5.201 {\footnotesize $\times 10^{-3}$} &	0.1334 &1.034 {\footnotesize $\times 10^{-3}$} \\	
Mean & 0.2487  & 5.541  {\footnotesize $\times 10^{-3}$} &0.1419 & 1.009 {\footnotesize $\times 10^{-3}$} \\
 Standard deviation & 0.0171 &  0.889 {\footnotesize $\times 10^{-3}$} & 0.0319 & 0.085{\footnotesize $\times 10^{-3}$} \\
\hline		
\hline		
\end{tabular}
\caption{Pertinent statistics of the parameter posterior pdfs of time-invariant parameters in Figure \ref{fig:case02_pdf}.}
\label{table:2}
\end{table}

\begin{figure}[H]
\centering
\begin{subfigure}{\linewidth}
\centering
\includegraphics[width=0.8\linewidth,keepaspectratio]{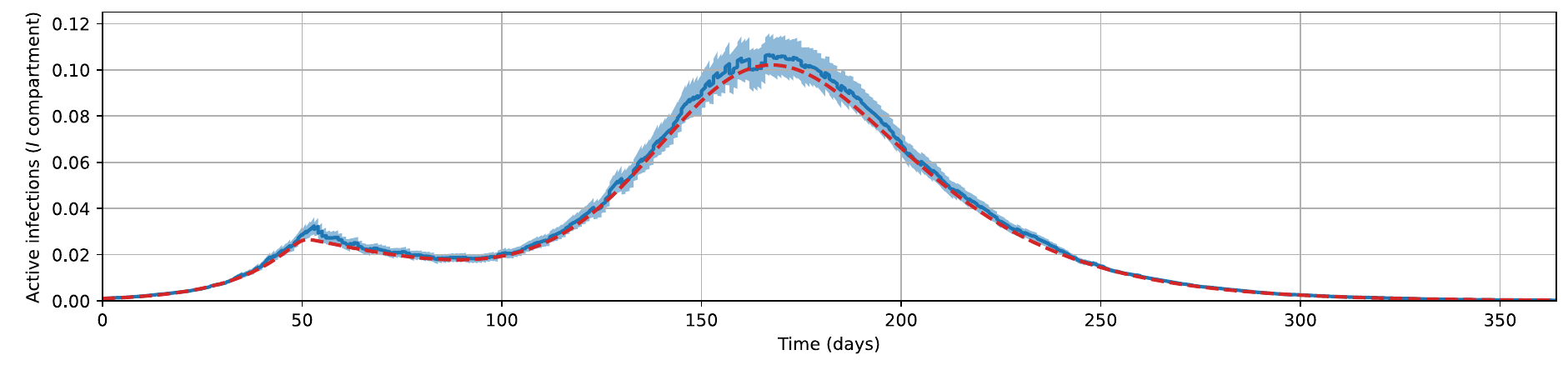}
\caption{}\label{fig:case02_i}
\end{subfigure}
\\
\begin{subfigure}{\linewidth}
\centering
\includegraphics[width=0.8\linewidth,keepaspectratio]{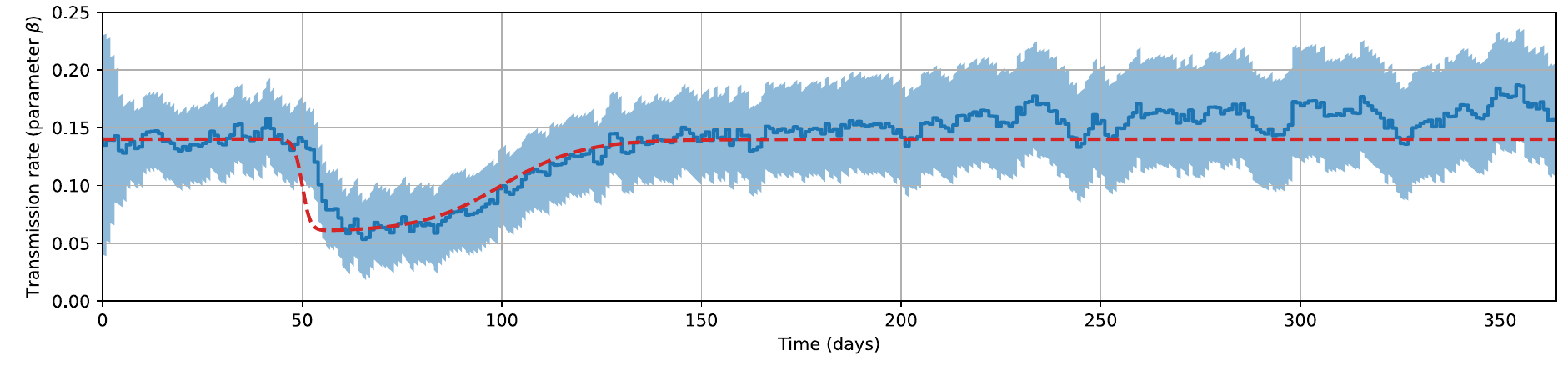}
\caption{}\label{fig:case02_beta}
\end{subfigure}
\\
\begin{subfigure}{\linewidth}
\centering
\includegraphics[width=0.8\linewidth,keepaspectratio]{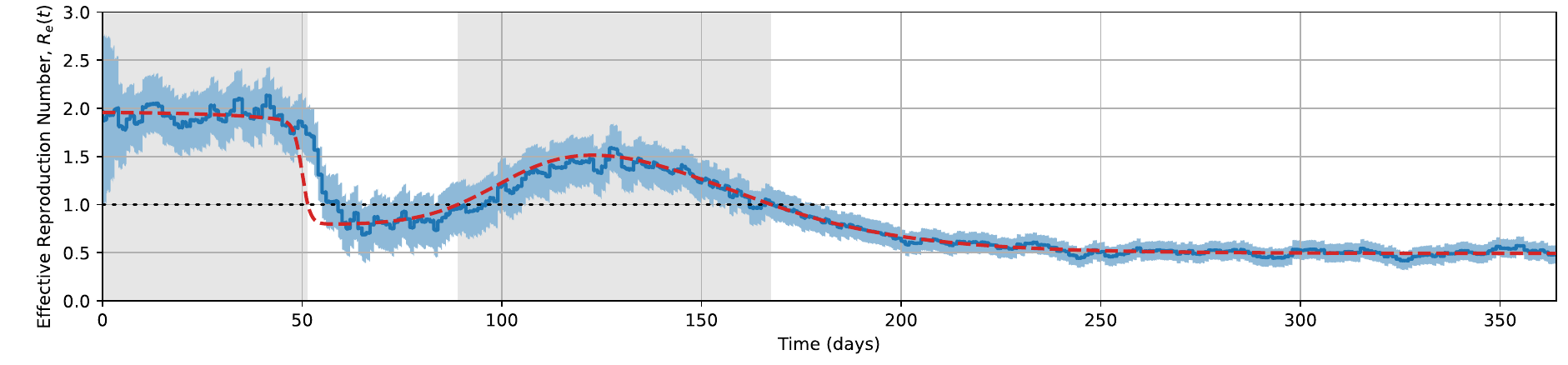}
\caption{}\label{fig:case02_re}
\end{subfigure}
\caption{State estimation results $\text{p}(\mathbf{x}_1,\hdots,\mathbf{x}_K \vert \mathbf{D}, \bm{\Phi}_s^{\text{MAP}},\mathcal{M})$ for (a) the infectious compartment, $I$, (b) the time-varying infection rate parameter, $\beta$, and (c) the effective reproduction number $R_e$. The ground truth is indicated by a dashed line ($\textcolor{tabred}{\blacksquare}$), the mean estimates are indicated by a solid line ($\textcolor{tabblue}{\blacksquare}$), and the shaded area reflects the mean $\pm$ 3 standard deviations.}\label{fig:case02_state}
\end{figure}

For the same reasons explained in the case of the seasonally varying infection rate parameter, the estimates of $\beta(t)$ appear to degrade as time progresses in panel \ref{fig:case02_beta}. The partially observed infectious compartment, and estimated effective reproduction number are both tracked well in time in panels \ref{fig:case02_i} and \ref{fig:case02_re}.

\subsection{Public health data: Ontario (March 1, 2020 - February 28, 2021)} \label{sec:ontario}
The province of Ontario publishes a number of COVID-19 metric in their public \textit{Data Catalogue}. The data used in this example is the reported number of active cases, from the column labelled \textit{confirmed positive} in the data set \cite{OntarioI}. Cases that are classified as active, is based on an assumed infectious period of 14 days. Thus, a case is considered positive for 14 days following the reported date of symptom onset or estimated episode date. We have selected one-year worth of testing data, collected between April 1, 2020 and March 31, 2021, such that the start date of our inference procedure aligns with the date this definition of an active case was adopted. Similarly, the selected end date aligns with the more widespread availability of vaccines in the province, thus pharmaceutical interventions have a minimal effect on the observed trends. Though the province of Ontario administered its first vaccines on December 14, 2020, the widespread distribution of vaccines to the general population did not begin until after the one-year period we consider in this case study. Hence, we intentionally omit the explicit modelling of the vaccination, and include any effect the early doses had on the population-level metrics to be absorbed by the time-varying infection rate parameter.

Each new case will be considered active for the average duration of infection, given by the inverse of the recovery rate parameter $\gamma$. As noted in Table \ref{table:priors}, the rate of recovery corresponds to an infectious period of 14 days. 

\begin{figure}[!htb]
\centering
\includegraphics[width=0.45\textwidth]{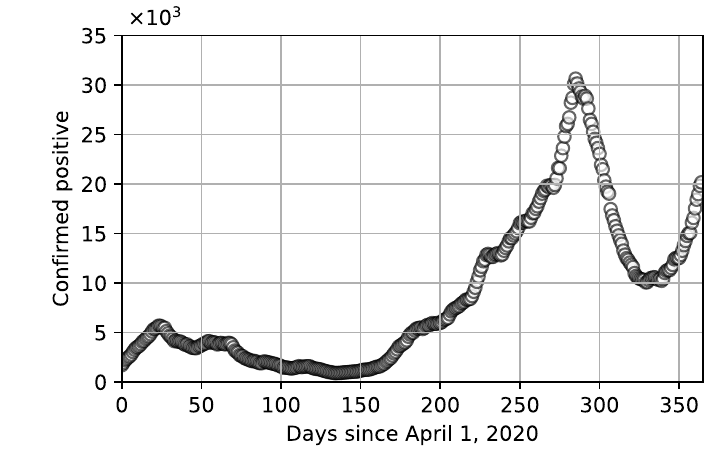}
\caption{Ontario public health data reporting active cases for the 365 days from April 1, 2020 to March 31, 2021.}
\label{fig:real_data}
\end{figure}

\begin{figure}[!htb]
\centering
\includegraphics[width=0.8\textwidth]{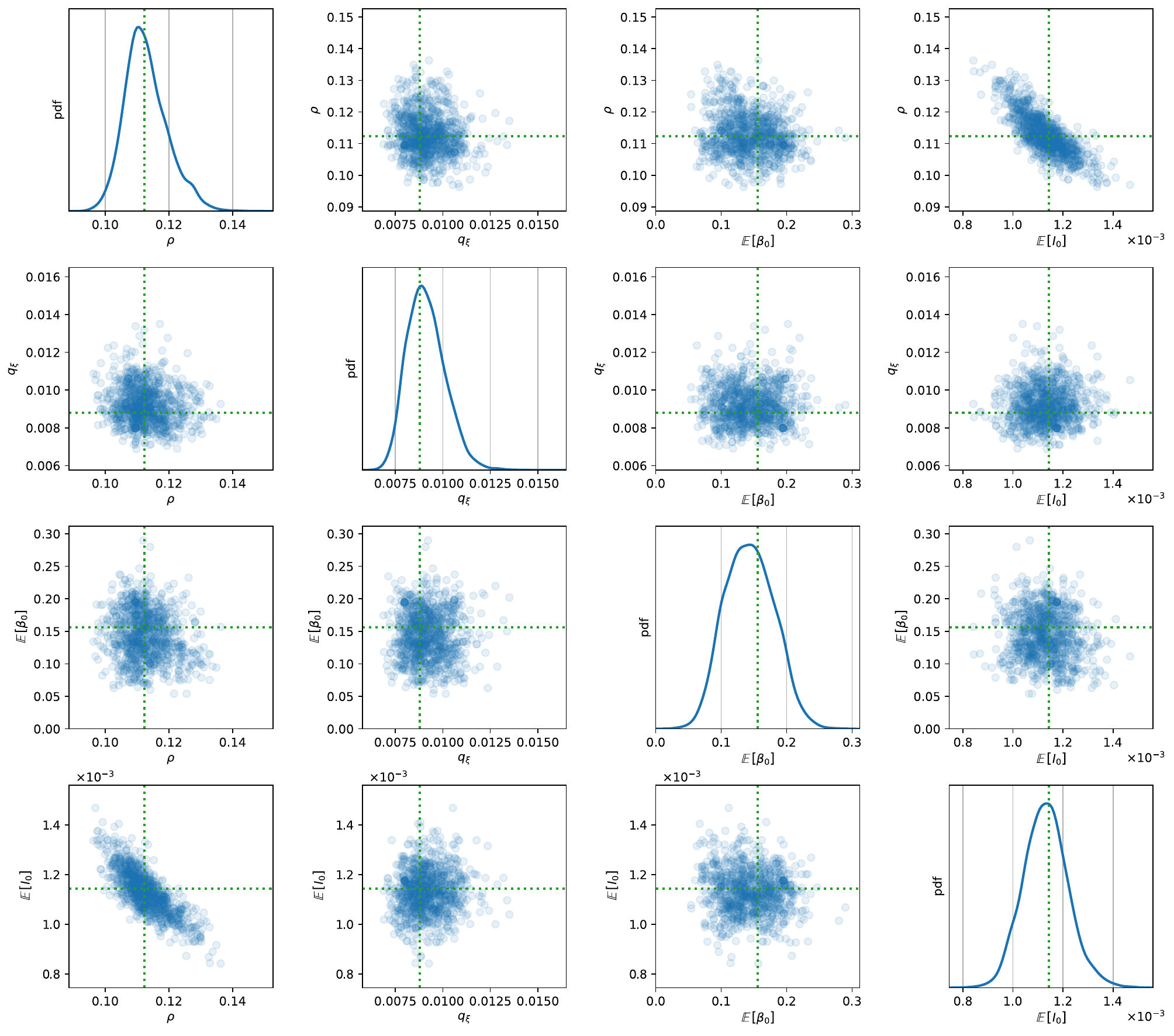}
\caption{Marginal parameter posterior pdfs and scatterplots of pairwise joint parameter posterior pdfs. MAP estimates are indicated by dotted lines ($\textcolor{tabgreen}{\blacksquare}$).}
\label{fig:real_pdfs}
\end{figure}

Through the examples with synthetic data, we demonstrated the benefits of jointly estimating the state, time-varying parameters using nonlinear filters, and the time-invariant parameters using MCMC. Figure \ref{fig:real_pdfs} shows the marginal and pairwise joint posterior pdfs of the time-invariant parameters. As in the case of synthetic data (Figures \ref{fig:case00_pdf}, \ref{fig:case01_pdf}, and \ref{fig:case02_pdf}), the mean initial estimate $\mathbb{E}[I_0]$ and the fraction of observed infections show substantial correlation.

\begin{figure}[!htb]
\centering
\begin{subfigure}{\linewidth}
\centering
\includegraphics[width=0.8\linewidth,keepaspectratio]{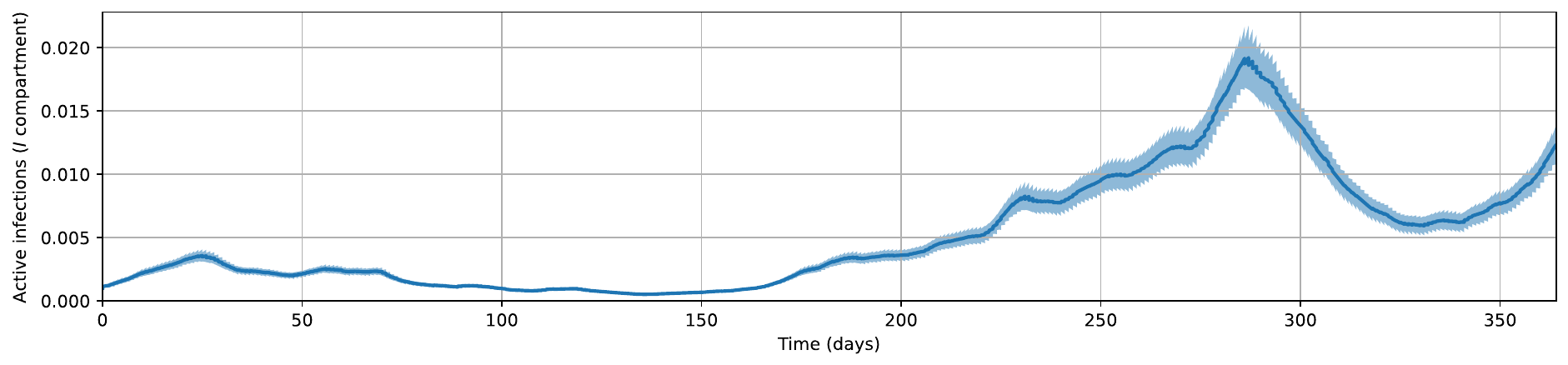}
\caption{}\label{fig:i_estimates}
\end{subfigure}
\\
\begin{subfigure}{\linewidth}
\centering
\includegraphics[width=0.8\linewidth,keepaspectratio]{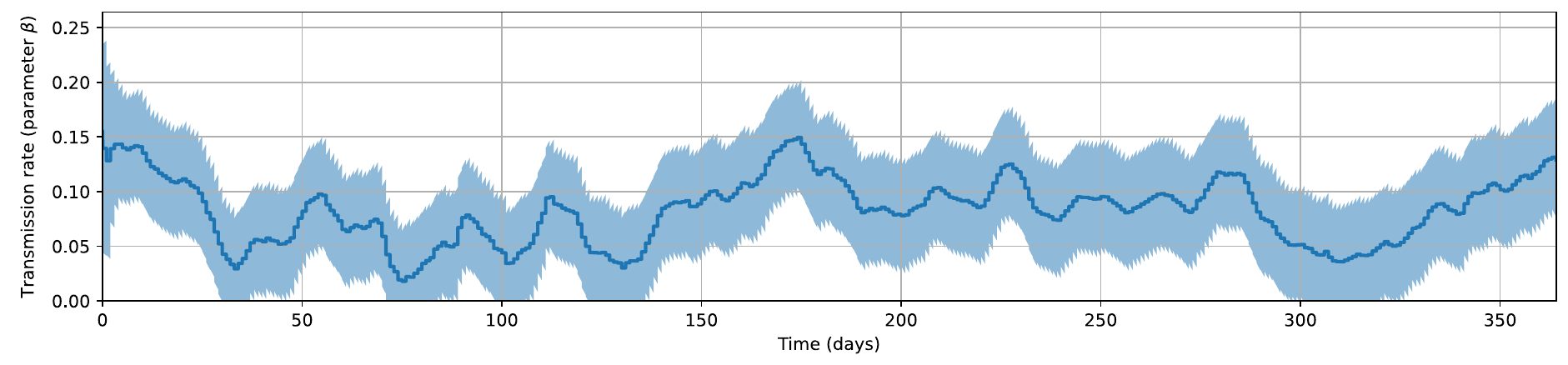}
\caption{}\label{fig:beta_estimates}
\end{subfigure}
\\
\begin{subfigure}{\linewidth}
\centering
\includegraphics[width=0.8\linewidth,keepaspectratio]{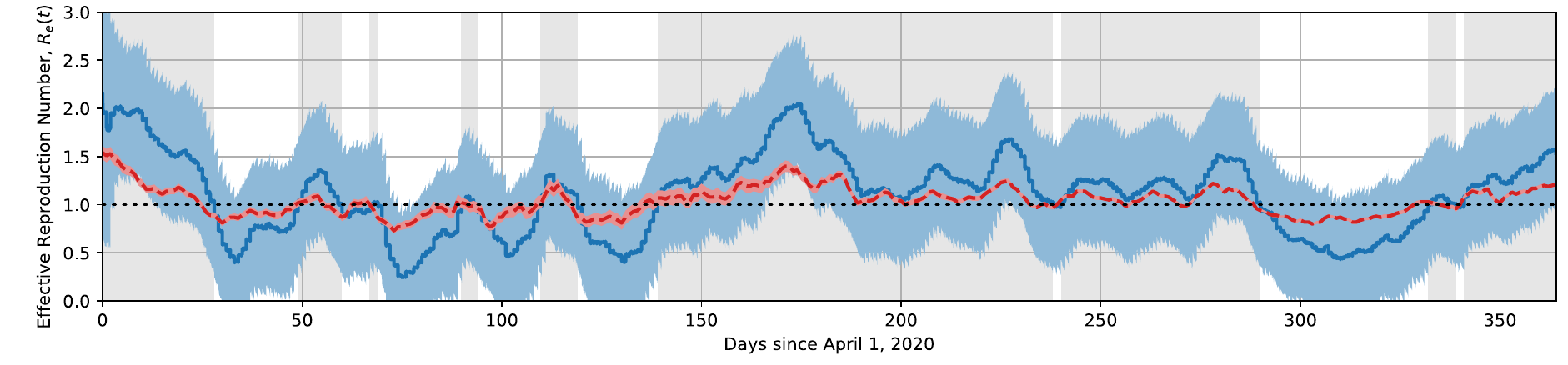}
\caption{}\label{fig:re_estimates}
\end{subfigure}
\caption{Our effective reproduction number estimate is indicated by a solid line ($\textcolor{tabblue}{\blacksquare}$), and the Ontario Public Health estimates are shown in a dashed line ($\textcolor{tabred}{\blacksquare}$) for 365 days from April 1, 2020 to March 31, 2021. The shaded area for our estimate reflects the mean $\pm$ 3 standard deviations, and the shaded area from Ontario Public Health in orange reflects the reported upper and lower bound confidence intervals. A dotted line ($\blacksquare$) is  included at $R_e = 1$, to assist in comparing the estimated timing of outbreaks.}\label{fig:ONresults}
\end{figure}

For the case with public health data, there is no known truth to provide a basis for comparison.  Hence, instead, we provide the estimated effective reproduction number $R_e$ in Figure \ref{fig:re_estimates} based on the joint estimates of the susceptible compartment, $S$, and time-varying infection rate parameter, $\beta$ as in Eq. (\ref{eq:Re}). We have superimposed the the estimates published by Ontario Public Health \cite{OntarioRe}. At first glance the two sets of results appear to be quite disparate. However, comparing the two sets of results relative to an effective reproduction number of $R_e = 1$ (indicated by a dotted in Figure \ref{fig:re_estimates}) the observed trends between the means of the two results are actually qualitatively quite similar in shape. The mean prediction of our results suggest that the reproduction number varies much more dramatically in time, compared to the published results which instead suggest the reproduction number hovered around 1 over the course of the one-year period studied here. Critically, note the difference in the reported confidence interval for the published results, compared to the uncertainty bounds obtained by in the current study. The three-standard deviation uncertainty bounds which envelopes the reported mean prediction almost fully captures the effective reproduction number published by Ontario Public Health . Given the modelling simplifications that are imposed when modelling a global phenomenon using compartmental models, and uncertainties in the system states and parameters, it is critical that modelling approaches adequately reflect the degree of uncertainty in the predictions, particularly when they form a basis for decision-making. 

\subsubsection{Forecast using public health data}
\begin{figure}[!htb]
\centering
\begin{subfigure}{0.33\linewidth}
\centering
\includegraphics[width=\linewidth,keepaspectratio]{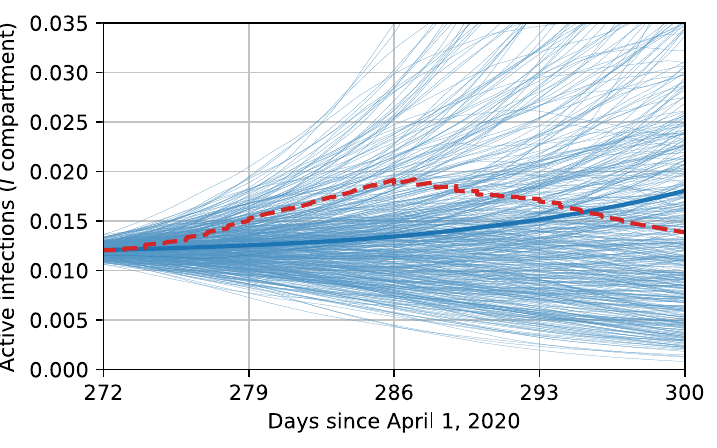}
\caption{}\label{fig:forecast_i1}
\end{subfigure}
\begin{subfigure}{0.33\linewidth}
\centering
\includegraphics[width=\linewidth,keepaspectratio]{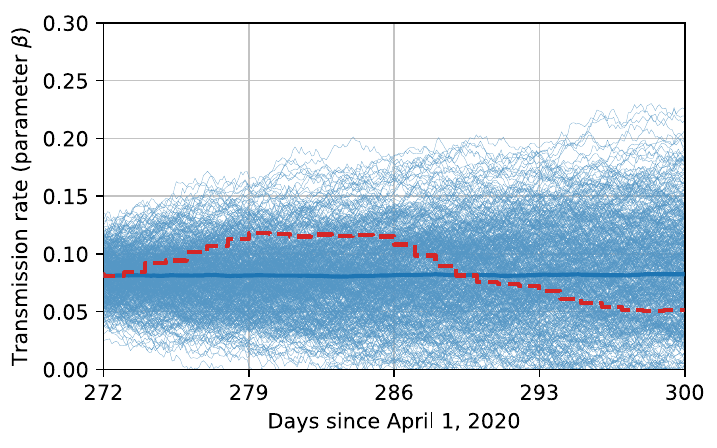}
\caption{}\label{fig:forecast_b1}
\end{subfigure}
\\
\begin{subfigure}{0.33\linewidth}
\centering
\includegraphics[width=\linewidth,keepaspectratio]{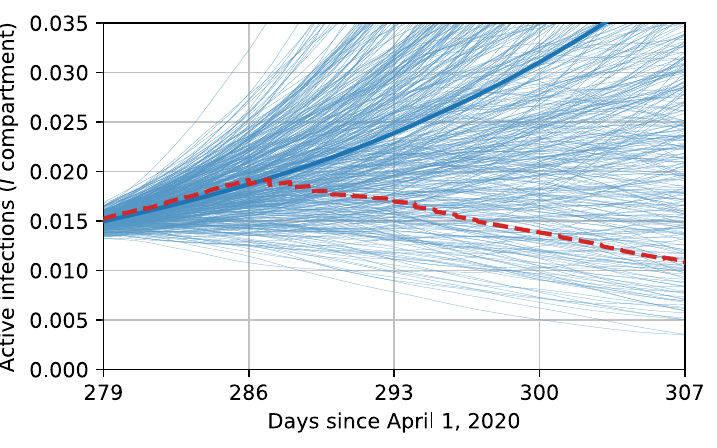}
\caption{}\label{fig:forecast_i2}
\end{subfigure}
\begin{subfigure}{0.33\linewidth}
\centering
\includegraphics[width=\linewidth,keepaspectratio]{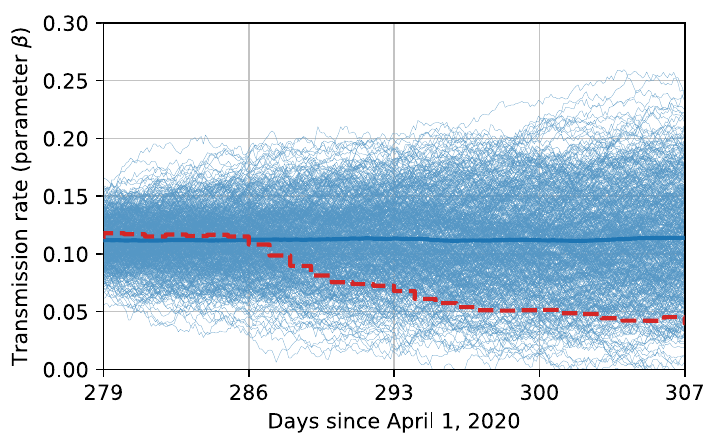}
\caption{}\label{fig:forecast_b2}
\end{subfigure}
\\
\begin{subfigure}{0.33\linewidth}
\centering
\includegraphics[width=\linewidth,keepaspectratio]{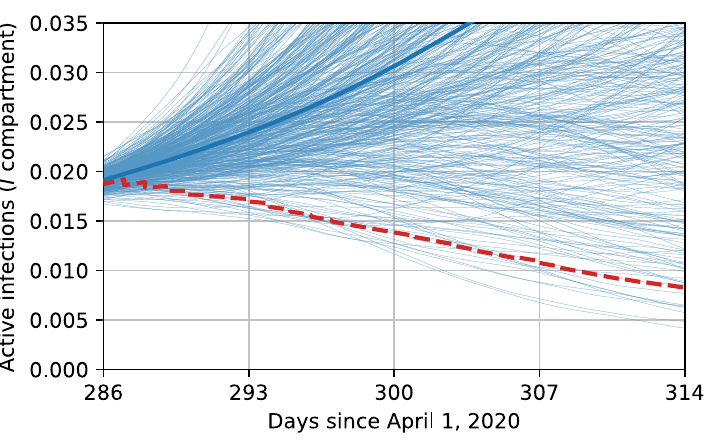}
\caption{}\label{fig:forecast_i3}
\end{subfigure}
\begin{subfigure}{0.33\linewidth}
\centering
\includegraphics[width=\linewidth,keepaspectratio]{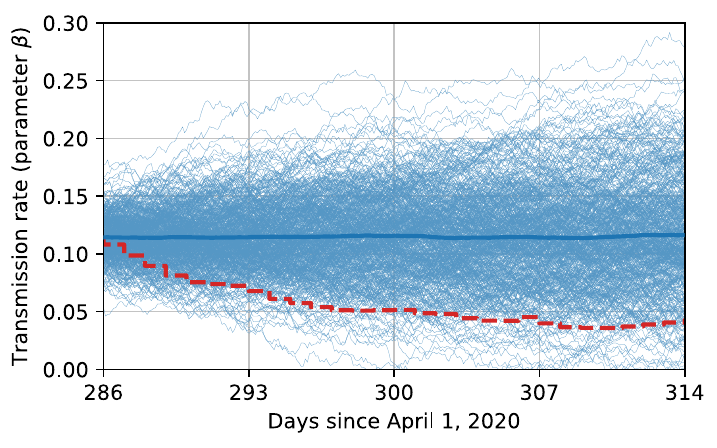}
\caption{}\label{fig:forecast_b3}
\end{subfigure}
\\
\caption{Three four-week forecasts in the vicinity of the peak of the infection curve, occurring on day 286. The forecasts begin (a) and (b) two weeks prior to the peak on day 272, (c) and (d) one week prior to the peak on day 279, and (e) and (f) at the peak on day 286. The estimates that were obtained by continuously assimilating data are indicated by a dashed line ($\textcolor{tabred}{\blacksquare}$). The forecasts are illustrated using 500 Monte Carlo samples with the mean estimate indicated by a solid line ($\textcolor{tabblue}{\blacksquare}$).}\label{fig:forecastON}
\end{figure}

Similar to Section \ref{synthetic_random} involving synthetic data,  we now present three four-week forecasts in Figure \ref{fig:forecastON}. These forecasts are obtained by integrating full nonlinear SIR model using Monte Carlo simulations, whereby the initial condition for each element of the augmented state is generated from a four-dimensional Gaussian distribution, whose mean vector and covariance matrix are given by the analysis step in the state estimation procedure (see Appendix \ref{sec:appendixA}) coinciding with the last available data point. In reference to the estimation results presented in Figure \ref{fig:ONresults}, these forecasts are performed in the vicinity of the peak of the infection curve, which occurs on day 286 (corresponding to January 12, 2021).  As the ground truth in this scenario is ultimately unknown, we compare the forecast results beyond the cutoff dates to the mean estimate of the infectious compartment that is obtained by assimilating data every day. For all three cases, it can be noted that the ensemble of Monte Carlo sample trajectories capture the future trajectory. 

Referring to  Figure \ref{fig:ONresults}, it can be noted that at day 272, the value of $\beta(t)$ (and thus, $R_e(t)$) increases suddenly for a week until day 279 where the effective reproduction number reaches a plateau and remains relatively constant for a week before decreasing sharply at day 286. The sudden increase in $\beta(t)$ following the last data point at day 272 seen in panel \ref{fig:forecast_b1}, causes the mean forecast in panel \ref{fig:forecast_i1} to underestimate the future trajectory. Conversely the sharp decrease in $\beta(t)$ following day 286 in panel \ref{fig:forecast_b3} results in an overestimate in panel \ref{fig:forecast_i3}. By this logic, it is understood that the relatively constant value of $\beta(t)$ between days 279 and 286 in panel \ref{fig:forecast_b2} results in good agreement between the forecast and the future trajectory over that same period in panel \ref{fig:forecast_i2}. 

It should be noted that the nature of the mean estimate in these cases is a reflection of the chosen model of the infection rate parameter $\beta(t)$ in Eq. (\ref{eq:randombeta}), rather than of the Bayesian computational framework. The events leading up to the peak at day 286 included the holiday season in Ontario and a subsequent province-wide shutdown. These are two significant events that would have heavily influenced the number of COVID-19 cases reported over this period of time, yet the model does not account for either of these explicitly in the model of $\beta(t)$. Nevertheless, through the joint estimation of the model parameters and the augmented state matrix leading up to these events,  the framework itself captured the significant rise and subsequent fall in cases as possible (if unlikely) trajectories.

\section{Conclusion}
Though we demonstrate the performance of the Bayesian computational framework for COVID-19 inspired problems, the framework itself is general. It is portable to other compartmental model structures, and more broadly to other models of dynamical systems. The generality of the framework makes it useful for both retrospective analyses and studies for post-pandemic recovery, as well as a versatile for future pandemic preparedness. Through three illustrative examples using synthetic data, the methodology proved to be adept at recovering the time-invariant and time-varying data-generating model parameters. For the case where the infection rate parameter was modelled by a Wiener process, the algorithm successfully estimated the strength of the artificial noise jointly with the parameter controlling the proportion of detected cases, as well as the mean initial conditions for the augmented state vector. For the cases where the infection rate instead had a known functional form, there is no ground truth value for the artificial noise strength. The estimation procedure bypasses the need for manual tuning of the artificial noise strength, by including this parameter among the time-invariant parameters and the Bayesian framework automatically favours the lowest noise strength that permits the filter to track the observed signal optimally. Even for a simple SIR model, the generality of this method relaxes many modelling assumptions, particularly those related to the initial conditions, which are of critical importance. Applying this validated methodology to public health data, the algorithm was shown to produce qualitatively similar results to those published by public health agencies. For instance, estimating the ratio of observed infections to active infections being at 11.2\% with a standard deviation of 0.6\%. The estimating this quantity as a time-invariant parameter through the framework leads to the estimates of the basic reproduction number differing in variability relative to the values reported by the agency. 

\bibliographystyle{unsrt}
{\footnotesize
\bibliography{refs}
}
\numberwithin{equation}{section}
\renewcommand\thefigure{C.\arabic{figure}}    
\setcounter{figure}{0}   
\setcounter{equation}{0}   


\begin{appendices}
\section{State Estimation}\label{sec:appendixA}
In this application, the state estimation procedure serves two purposes; to estimate the augmented state of the system (compartments and time-varying parameter) given the proposed model, the set of static parameters, and the incoming noisy sensor data, and to compute the likelihood function for the joint estimation of the time-invariant parameters. As we augment the state to include the time varying parameters, we have now introduced additional nonlinearity into the state space model, and hence, nonlinear filters are required for state estimation. Here we use the extended Kalman filter (EKF) for all cases, noting that this imposes the assumption that the state pdfs are approximately Gaussian, which generally holds for weakly nonlinear systems, complemented by dense/frequent observations \cite{khalil2009nonlinear}. 

The EKF imposes a Gaussian assumption on the state such that $\mathbf{x}_k \sim \mathcal{N}(\mathbf{x}_k^a,\mathbf{P}_k^a)$, where $\mathbf{x}_k^a$ and $\mathbf{P}_k^a$ are the update (or analysis) mean vector and covariance matrix, respectively. Similarly, the process noise is assumed to be normally distributed with $\mathbf{q}_k \sim \mathcal{N}(\mathbf{0},\mathbf{Q}_k)$. For points along the computational grid, indexed by $k$, that do not coincide with a data point, indexed by $d(j)$, the current mean state estimate and its uncertainty are forecasted according to \cite{jazwinski2007stochastic}

\begin{align}
\mathbf{x}_{k+1}^f&=\mathbf{g}_k(\mathbf{x}_k,\bm{\Phi}_s, \mathbf{0}), \label{eq:forecast_x} \\
\mathbf{P}_{k+1}^f&= \mathbf{A}_k\mathbf{P}_k^a\mathbf{A}_k^T + \mathbf{B}_k\mathbf{Q}_k\mathbf{B}_k^T. \label{eq:forecast_P}
\end{align}
where the Jacobian matrices

\begin{align}
\mathbf{A}_k &= \left. \frac{\partial \mathbf{g}_{k}(\mathbf{x}_k^a,\bm{\Phi}_s, \mathbf{q}_k)}{\partial \mathbf{x}_{k}} \right\vert_{\mathbf{x}_k = \mathbf{x}_{k}^a, \mathbf{q}_k = \mathbf{0}} \label{eq:ekf_update1}, \\
\mathbf{B}_k &=  \left. \frac{\partial \mathbf{g}_{k}(\mathbf{x}_{k},\bm{\Phi}_s, \mathbf{q}_k)}{\partial \mathbf{q}_k} \right\vert_{\mathbf{x}_k = \mathbf{x}_k^a, \mathbf{q}_k = \mathbf{0}}  \label{eq:ekf_update2}.
\end{align}

When the computational grid coincides with a measurement, the analysis step in Eqs. (\ref{eq:a5} - \ref{eq:a7}), is performed, allowing for the mean and covariance of the state estimate to be updated. The measurement operator is given in Eq. (\ref{eq:meas}), where the measurement noise is assumed to be Gaussian $\varepsilon_j \sim \mathcal{N}(\mathbf{0},\bm{\Gamma}_j)$, where $\Gamma_j$ is the covariance matrix, resulting in the following equations \cite{jazwinski2007stochastic} 

\begin{align}
\mathbf{x}_{d(j)}^{a}&=\mathbf{x}_{d(j)}^f + \mathbf{K}_{d(j)}(\mathbf{d}_j - \mathbf{h}_{d(j)}(\mathbf{x}_{d(j)}^f,\mathbf{0})), \label{eq:a5} \\
\mathbf{P}_{d(j)}^a&= (\mathbf{I} - \mathbf{K}_j\mathbf{C}_j)\mathbf{P}_{d(j)}^f, \label{eq:a6} \\
\mathbf{K}_{d(j)}&= \mathbf{P}_{d(j)}^f \mathbf{C}_j^T \left[\mathbf{C}_j \mathbf{P}_{d(j)}^f \mathbf{C}_j^T + \mathbf{D}_j \bm{\Gamma}_{j} \mathbf{D}_j^T  \right]^{-1},  \label{eq:a7}
\end{align}
where the Jacobian matrices $\mathbf{C}_j $ and $\mathbf{D}_j$ are evaluated as

\begin{align}
\mathbf{C}_j &= \left. \frac{\partial \mathbf{h}_{d(j)}(\mathbf{x}_{d(j)}, \bm{\varepsilon}_j)}{\partial \mathbf{x}_{d(j)}} \right\vert_{\mathbf{x}_{d(j)} = \mathbf{x}_{d(j)}^f, \mathbf{\varepsilon}_{j} = \mathbf{0}} \label{eq:ekf_update1}, \\
\mathbf{D}_j &=  \left. \frac{\partial \mathbf{h}_{d(j)}(\mathbf{x}_{d(j)}, \bm{\varepsilon}_j)}{\partial \bm{\varepsilon}_j} \right\vert_{\mathbf{x}_{d(j)} = \mathbf{x}_{d(j)}^f, \mathbf{\varepsilon}_j = \mathbf{0}}  \label{eq:ekf_update2}.
\end{align}

Critically, when data are available, we can also compute the likelihood function in Eq. (\ref{eq:lik}) according to \cite{bisaillon2015bayesian,khalil2009nonlinear,khalil2015estimation,khalil2013probabilistic}

\begin{align}
 \text{p}(\mathbf{d}_j\vert \bm{\Phi}_s,\mathcal{M}_i) =  \mathcal{N}\left(\mathbf{d}_j | \mathbf{h}_{j}\left(\mathbf{x}_{d(j)}^f, \mathbf{0}\right), \bm{\Sigma '}\right)  \label{eq:int_ekf}
\end{align}
where 
\begin{equation}
\bm{\Sigma '} = \mathbf{C}_j \mathbf{P}_{d(j)}^f \mathbf{C}_j^T + \mathbf{D}_j \bm{\Gamma}_j \mathbf{D}_j^T
\end{equation}

\section{Jacobian Matrices}\label{sec:appendixB}
The model operator in Eq. (\ref{eq:sirb}) can be written in terms of the elements of the state vector $\mathbf{x}_k$,

\begin{subequations}\label{eq:xirb}
\begin{align}
x_{1,k+1} &= x_{1,k} + \Delta t \left( -x_{1,k} x_{2,k} x_{4,k}  \right), 
\\
x_{2,k+1} &= x_{2,k} + \Delta t \left( x_{1,k} x_{2,k} x_{4,k} - \gamma x_{2,k} \right), 
\\
x_{3,k+1} &= x_{3,k} + \Delta t \left( \gamma x_{2,k} \right),
\\
x_{4,k+1} &= x_{4,k} + q_\xi \sqrt{\Delta t}  \xi_{k}.
\end{align}
\end{subequations}

The above equations lead to the following expressions for the Jacobian matrices,

\begin{equation}
\mathbf{A}_k = { \left[ \begin{array}{cccc}
1 + \Delta t \left( - x_{2,k} x_{4,k}  \right) 
&
\Delta t \left( -x_{1,k}x_{4,k}  \right)
&
0
&
\Delta t \left( -x_{1,k} x_{2,k}   \right)
\\
\Delta t \left( x_{2,k} x_{4,k}  \right)
&
1 + \Delta t \left( x_{1,k} x_{4,k} - \gamma \right)
&
0
&
 \Delta t \left( x_{1,k} x_{2,k} \right)
\\
0
&
\Delta t \left( \gamma \right)
&
1 
&
0
\\
0
& 
0
&
0
&
1
\end{array}\right]}, 
\qquad 
\mathbf{B}_k =  { \left[ \begin{array}{ccccc}
0 \\ 0 \\ 0 \\ q_\xi \sqrt{\Delta t}
\end{array}\right]}
\end{equation}

The measurement operator represents a partial measurement of the infectious compartment. The ratio of detected to undetected infections is reflected in the parameter $\rho$, 
\begin{equation}
d_j = \rho x_{2,j}\left(1 + \epsilon_j \right)
\end{equation}

The Jacobian matrices of the measurement operator are

\begin{equation}
\mathbf{C}_j =  { \left[ \begin{array}{ccccc}
0&1 & 0 &  0
\end{array}\right]},  
\qquad 
\mathbf{D}_j = \left[ \begin{array}{c}
x_{2,d(j)}
\end{array}\right]
\end{equation}

\end{appendices}

\bibliographystyle{unsrt}  
\bibliography{refs}  

\end{document}